\documentclass[smallextended]{svjour3wide} 
\smartqed  
\usepackage[dvipdfmx]{graphicx}
\usepackage{subfigure}
\usepackage{amsmath}
\usepackage{amssymb}
\usepackage{mathrsfs}
\usepackage{color}
\usepackage{bm}
\usepackage{ulem}
\usepackage{here}
\usepackage{setspace}

\journalname{Journal of Statistical Physics}
\begin{document}

\title{Phase growth with heat diffusion in a stochastic lattice model}

\author{Mao Hiraizumi \and  Hiroki Ohta \and Shin-ichi Sasa}

\institute{
M. Hiraizumi \at
Department of Physics, Kyoto University, Kyoto 606-8502, Japan \\
\email{hiraizumi.mao.72s@st.kyoto-u.ac.jp}
\and H. Ohta \at Department of Human Sciences, Obihiro University of Agriculture and Veterinary Medicine, Hokkaido 080-8555, Japan\\ \email{hirokiohta@obihiro.ac.jp}
\and S. Sasa \at  Department of Physics, Kyoto University, Kyoto 606-8502, Japan \\
\email{sasa@scphys.kyoto-u.ac.jp}
}

\date{Received: date / Accepted: date}

\maketitle

\begin{abstract}
 When a stable phase is adjacent to a metastable phase with a planar interface,
the stable phase grows. We propose a stochastic lattice model describing
the phase growth accompanying heat diffusion. The model is based on an
energy-conserving Potts model with a kinetic energy term defined
on a two-dimensional lattice, where each site is sparse-randomly
  connected in one direction and local in the other direction.
For this model,
we calculate the stable and metastable phases exactly using
statistical mechanics.  Performing numerical simulations, 
we measure the displacement of the interface $R(t)$.
We observe the scaling relation $R(t)=L_x \bar{\mathcal{R}} (Dt/L_x^2)$, where $D$ is the thermal diffusion constant and $L_x$ is the system size between the two heat baths. The scaling function $\bar{\mathcal{R}}(z)$ shows $\bar{\mathcal{R}}(z) \simeq z^{0.5}$ for $z \ll z_c$ and $\bar{\mathcal{R}}(z) \simeq z^{\alpha}$ for $z \gg z_c$, where the cross-over value $z_c$ and exponent $\alpha$ depend on the temperatures of the baths, and $0.5\le\alpha\le 1$.
We then confirm that a deterministic phase-field model exhibits the
same scaling relation. Moreover, numerical simulations of the
phase-field model show that the cross-over
value $\bar{\mathcal{R}}(z_c)$ approaches zero when the stable
phase becomes neutral.
\end{abstract}

\section{Introduction}


When a stable phase contacts a metastable phase, the stable phase grows and  the metastable phase eventually vanishes. This phenomenon is ubiquitously observed in nature \cite{Callen}. The basic understanding of phase growth is that the propagation velocity of the interface between the two phases is proportional to the difference in free energy densities, where the proportional constant is the mobility \cite{Pomeau}. This mechanism is applied to isothermal systems where the energy locally dissipates in a heat bath.
However, qualitatively different behavior is observed in systems where the energy is locally conserved. In such a system, 
latent heat generated at the interface in a growth process induces a change in local temperature of the interface, and the energy transfers
  into the bulk region through heat diffusion. This effect
modifies the law of interface propagation.


There are several types of deterministic models for interface motion with heat diffusion. The most
classical is the heat diffusion model with moving boundary condition
at the interface, which was formulated by Stefan \cite{Stefan1,Stefan2}.
In this description, called the sharp-interface model, the 
interface is regarded as a singular region of the continuous temperature
field. Although this is mathematically defined, it is not easy to calculate
the interface motion numerically. A computationally efficient model
that takes heat diffusion into account is the phase-field model \cite{fix,Cagphase,Langphase,CL,penrose,kobayashi}.
This model is given as a set of coupled partial differential equations of the order parameter field and the temperature field. 
Both models show that 
the displacement of the planar interface is proportional to the square root of the time interval 
 when the extent of the metastability $\Delta$ is less than unity \cite{Zener,Dewynne,Lowen,HS}. Here, $\Delta$ is defined as
\begin{eqnarray} 
\Delta \equiv \frac{c_p}{T_c (\delta s)}|T_c-T_{\rm ms}|,
\label{Def_Delta}
\end{eqnarray}
where $T_c$ is the equilibrium transition temperature, $T_{\rm ms}$ is the temperature of the heat bath in contact with the metastable phase, $\delta s$ is the entropy jump per unit volume, and $c_p$ is the specific heat capacity per unit volume under constant pressure.


Now, the question we address in this paper is whether  thermal fluctuations influence the phase growth with heat diffusion.
According to non-equilibrium statistical mechanics, the starting point
of a mesoscopic description such as the phase-field model is the entropy
functional consisting of the spatial integration of the local entropy density
and the gradient term \cite{penrose,HHM,AMW}. The entropy density is a function of the internal energy density and the number density. Then, following the Onsager principle, one can determine the evolution equation of these densities so that the irreversible currents are given as linear combinations of thermodynamic forces. The obtained equation is equivalent to the phase-field model \cite{penrose}.
 This form of the phase-field model was
also introduced in the context of dynamical behavior near the critical
point \cite{HHM,HH}. 
Because such models derived from the Onsager principle are defined in a mesoscopic regime, thermal noises are inevitable in this description, where the noise intensity is determined by the
fluctuation-dissipation relation of the second kind. 
Even worse, the interface may be out of the
mesoscopic description, precisely speaking,  because the interface width is on the order of $10^{-7}$
cm \cite{TR}. We thus need to consider a more microscopic
model to study fluctuation effects. Although many statistical
mechanics models have been studied in the context of phase growth
\cite{BCF,GB,Muller,WGA,SaitoM,Gilmer,DLA,MH,NRKTR}, heat diffusion
has not been taken into account. For this reason, we propose a statistical
mechanics model describing the phase growth with heat diffusion.




The model we propose is the $q$-state Potts model \cite{Wu} with an additional variable representing the kinetic energy at each site, whose stochastic time evolution satisfies the detailed balance condition at equilibrium \cite{KS,Creutz,Casartelli}. A similar model without the kinetic energy term was
investigated to study ordering processes after quenching \cite{Sahni,KNG,SM}. In equilibrium statistical mechanics, we can determine the phase diagram at equilibrium. The model exhibits the order-disorder transition as the temperature is changed.  Including kinetic energy enables the model to describe the conversion from potential energy to kinetic energy, which corresponds
to the generation of latent heat. By introducing a transition rule with energy conservation, heat diffusion is described by kinetic energy exchanging processes. We note that the conventional Potts model without the kinetic term corresponds to the system where the generated latent heat is immediately dissipated into the heat bath.
We study the model defined on a two-dimensional lattice, where each site is sparse-randomly connected in one direction and local in the other direction. From this setting, we can precisely identify the metastable phase in addition to the equilibrium properties. 

We numerically simulate the model to measure the displacement of the interface $R(t)$. Let $T_{s}$ be the temperature of the heat bath in contact with the stable phase. We find the scaling relation $R(t)=L_x \bar{\mathcal{R}} (Dt/L_x^2)$, where $D$ is the thermal diffusion constant and  $L_x$ is the system size between the two heat baths. The scaling function $\bar{\mathcal{R}}(z)$ shows $\bar{\mathcal{R}}(z) \simeq z^{0.5}$ for $z \ll z_c$ and $\bar{\mathcal{R}}(z) \simeq z^{\alpha}$ for $z \gg z_c$, where the cross-over value $z_c$ and the exponent $\alpha$ depend on $T_s$, and $0.5\le\alpha\le 1$. Because the scaling relation in the late stage has not been reported in the phase-field model, the result could imply
that the stochastic phase growth involves a different universality class
from that described by the phase-field model. However, we have found that such a finite-size and long-time behavior is also observed in the phase-field model even without noise. This indicates that the phase-field model is more universal than that already known. Furthermore, systematic numerical simulations
of the phase-field model reveal the new feature that
the cross-over value $\bar{\mathcal{R}}(z_c)$ approaches zero
when the stable phase becomes neutral.

This paper is organized as follows. In Sec.~\ref{Model}, we introduce the model. In Sec.~\ref{SM}, we analyze the model via equilibrium statistical mechanics. We identify the metastable phase in addition to equilibrium properties. In Sec.~\ref{numerics}, we report the results of numerical simulations and compare them with the numerical results of the phase-field model.
We make some concluding remarks in Sec.~\ref{remark}. The technical details of the theoretical calculation are summarized in Appendix~\ref{app-sec:SM}. The values of $\Delta$ and $D$ are estimated in Appendix  \ref{app-sec:latent heat} and  Appendix \ref{app-sec:Diff}, respectively. Throughout the paper, the Boltzmann constant is set to unity, and $\beta$ is always connected to the temperature $T$ via $\beta=1/T$.

\section{Model}\label{Model}

\subsection{Hamiltonian}

\begin{figure}
\centering
    \subfigure[Square lattice.]{%
        \includegraphics[clip, width=0.4\columnwidth]{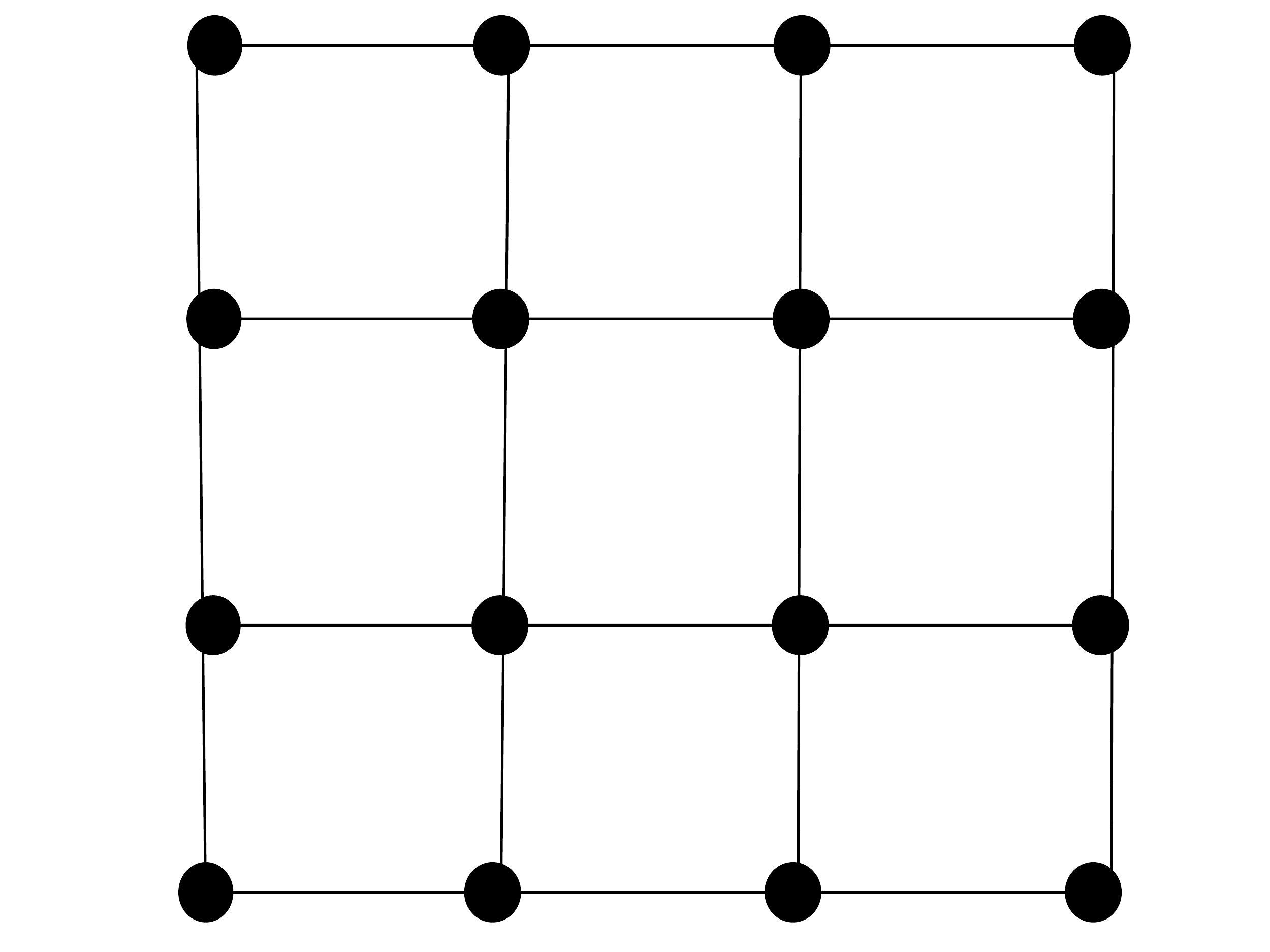}
        \label{schematic_1}
         }%
    \subfigure[Sparse-randomly connected lattice.]{%
        \includegraphics[clip, width=0.4\columnwidth]{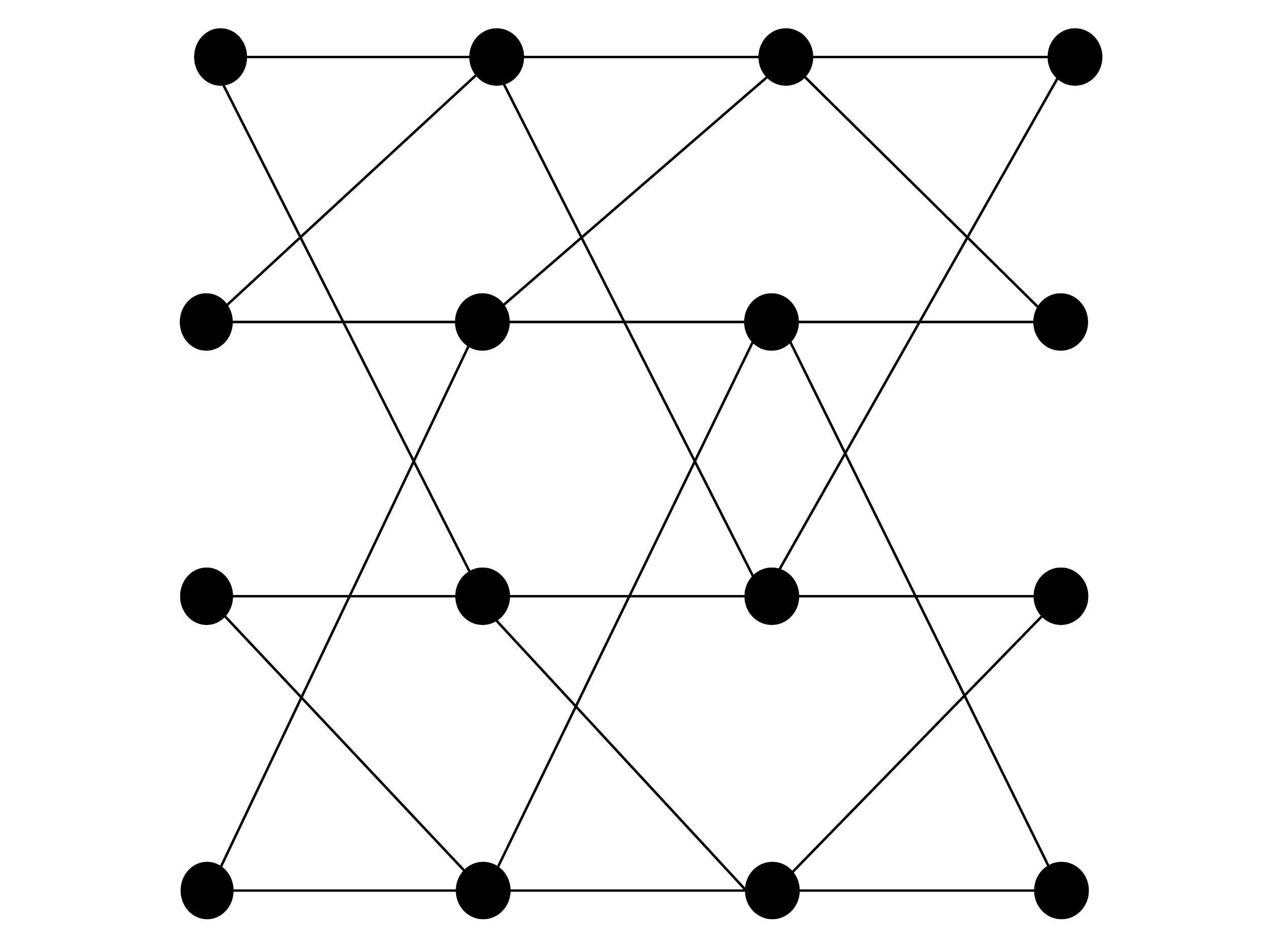}
      \label{schematic_2}
        }%
    \caption{Schematics of lattices.}
\end{figure}

Let $\Lambda =\{ i=(i_x,i_y) | 1\le i_x \le L_x, 
\quad 1\le i_y \le L_y,   i_x, i_y \in \mathbb{Z} \}$
be a two-dimensional lattice. For any site $i \in \Lambda$,
a collection of sites connected to site $i$ is denoted as
$B_i$. We assume that set $B_i$ is decomposed as
\begin{equation}
B_i=B_i^{-}\cup B_i^0 \cup B_i^+,
\end{equation}
where $j_x=i_x-1$ for $j \in B_i^{-}$, $j_x=i_x+1$ for $j \in B_i^{+}$,
and $j_x=i_x$ for $j \in B_i^{0}$. Note that $B_1^{-}=\emptyset$ and $B_{L_x}^+=\emptyset$. For example, 
$B_i^-=\{ (i_x-1,i_y) \}$, $B_i^+=\{ (i_x+1,i_y) \}$,
and $B_i^0=\{ (i_x,i_y\pm 1) \}$ for the square lattice in Fig.~\ref{schematic_1}. In this paper, we use a sparse-randomly connected lattice
defined by  $B_i^-=\{ (i_x-1,i_y), (i_x-1,b^-(i_x-1,i_y) \}$,
$B_i^+=\{ (i_x+1,i_y),(i_x+1,b^+(i_x,i_y)) \}$, 
and $B_i^0=\emptyset $, where $b^+(i_x,\ )$ is a one-to-one
random map from $\{1,\cdots,L_y \}$ to $\{1,\cdots,L_y \}$ that
satisfies $b^+(i_x,i_y)\not = i_y$, and $b^-(i_x,\ )$ is
defined by the inverse of the map. This is a special case of the random graphs introduced in \cite{ORT}. See Fig.~\ref{schematic_2} for the illustration.

On each site $i \in \Lambda$, the $q$-state variable
$\sigma_i\in \{1, 2,\cdots,q\}$ and the positive valued 
kinetic energy $p_i \in \mathbb{R}^{+}$ are defined. The collections
of variables, $[\sigma_i]_{i \in \Lambda}$ and $[p_i]_{i \in \Lambda}$, are
simply denoted by $\sigma$ and $p$. The Hamiltonian we study is 
\begin{equation}
H(\sigma,p)=-\sum_{i\in\Lambda}\frac{ 2 }{|B_i|} \sum_{j \in B_i}\delta(\sigma_i,\sigma_j)+\sum_{i\in\Lambda} p_i,
\end{equation}
where $\delta( \cdot ,\cdot )$ represents  the Kronecker delta.

\subsection{Model with energy conservation}
\label{rule_bulk}

Before presenting our model, we first describe a model for a thermally
isolated system. We describe the stochastic
time evolution in which the stationary distribution is given by
the microcanonical ensemble associated with the Hamiltonian
$H(\sigma,p)$. The stochastic process satisfies the detailed balance
condition with respect to the uniform distribution on the energy surface
$H(\sigma,p)=E$,  where $E$ is the total energy, which is invariant
under time evolution.

We perform the following five procedures at one step:
\begin{enumerate}
\item  A site $i \in \Lambda$ is chosen randomly with equal probability.

\item  The potential energy difference $dE$ is calculated for a transition
$\sigma_i \to \sigma_i'$ chosen randomly with equal probability. 

\item The transition $\sigma_i \to \sigma_i'$ is accepted if $p_i -dE \ge 0$,
and then the kinetic energy is changed to $ p_i'= p_i -dE$. 

\item A site $k$ and another site $j \in B_k$ are chosen randomly.

\item The transition $(p_{k},p_j) \to (p_{k}-dp, p_j+dp)$
 is accepted if $p_{k}-dp \ge  0$, where $dp$ is a numerical parameter.
 
\end{enumerate}
Let $(p_i^0)_{i \in \Lambda}$ be a collection of the initial value of 
kinetic energy for each site. From the evolution rule, 
$p_i$ is written as $p_i= p_i^{0}+n_i +m_i dp$,
where $n_i, m_i \in \mathbb{Z}$. The set
of all possible values of $p_i$ is denoted by $\mathbb{P}_i$.
The phase space of the model $\Sigma$ as a function of $E$
is then expressed by the discrete set 
\begin{eqnarray}
\Sigma(E)=\{(\sigma,p)|H(\sigma,p)=E, 
\ \sigma_i\in \{1,\cdots,q\},\ p_i \in \mathbb{P}_i,  \ i \in \Lambda\}.
\end{eqnarray}

The transition probability determined by the procedures is
expressed as $W(\sigma',p'|\sigma,p)$.
Let $t\in\mathbb{Z}$ be the discrete time and $P_t(\sigma,p)$ be the probability
of taking the state $(\sigma,p)$ at $t$-step. Then, we have
\begin{equation}
P_{t+1}(\sigma,p)=\sum_{\sigma',p'}W(\sigma,p|\sigma',p')P_t(\sigma',p').
\end{equation}
Because $W(\sigma',p'|\sigma,p)=W(\sigma,p|\sigma',p')$ holds,
the stationary distribution is given by the microcanonical
form
\begin{equation}
P_{\rm mc}(\sigma,p)= \frac{1}{|\Sigma(E)|}\delta(H(\sigma,p),E),
\label{ss-dis}
\end{equation}
where $|\Sigma(E)|$ denotes the number of elements of set $\Sigma(E)$.

\subsection{Model with two heat baths}
\label{rule_bc}

In the model, we attach two heat baths, one on the left side
$i_x=1$ and one on the right side $i_x=L_x$ of the system introduced in
the previous subsection. The temperatures of the left
and right heat baths are denoted by $T_L$ and $T_R$,
respectively. The stochastic time evolution of the model is
given by imposing an additional rule at the boundaries.
We perform  the following procedures
when a site $i$ with $i_x=1$ or $i_x=L_x$ is chosen in procedure 1 of the time evolution described in the previous subsection:
\begin{enumerate}
\item  The potential energy difference $d E$ is calculated for a transition
$\sigma_i \to \sigma_i'$ chosen randomly. 

\item The transition $\sigma_i \to \sigma_i'$ is accepted with the
probability $w(\sigma_i \to \sigma_i')$, where
\begin{align}
w(\sigma_i\to \sigma_i^\prime)=\frac{1}{2}
\left(1-\tanh\left(\frac{dE}{2T}\right)\right)
\label{w-def}
\end{align}
with $T=T_L$ for $i_x=1$ and $T=T_R$ for $i_x=L_x$.

\item The value of $p_i$ is replaced with a new one sampled
from the distribution 
\begin{align}
P(p_i)= \frac{1}{T}e^{-\frac{p_i}{T}}
\end{align}
with $T=T_L$ for $i_x=1$ and $T=T_R$ for $i_x=L_x$. We here note that $p_{i}$ takes positive value.
\end{enumerate}
Note that (\ref{w-def}) satisfies the detailed balance
condition 
\begin{align}
\frac{w(\sigma\to \sigma^\prime)}{w(\sigma^\prime\to \sigma)}=e^{-\frac{dE}{T}}.
\end{align}
The stochastic rule
in the previous subsection is used except at the boundaries $i_x=1$ and $i_x=L_x$. The transition probability determined by the procedures is
expressed as $\tilde W(\sigma',p'|\sigma,p)$.
Let $t\in\mathbb{Z}$ be the discrete time and
$\tilde P_t(\sigma,p)$ be the probability
of taking state $(\sigma,p)$ at each $t$-step. Then, we have
\begin{equation}
\tilde P_{t+1}(\sigma',p')=\sum_{\sigma,p}\tilde W(\sigma',p'|\sigma,p)
\tilde P_t(\sigma,p).
\end{equation}
When $T_L=T_R=T$, the detailed balance condition
\begin{align}
\frac{\tilde W(\sigma',p'|\sigma,p)}
{\tilde W(\sigma,p|\sigma',p')}=e^{-\frac{dE}{T}}
\end{align}
holds. Thus, the stationary distribution is given by the canonical form 
\begin{equation}
P_{\rm can}(\sigma,p)= \frac{1}{Z_{\rm tot}(\beta)}
e^{-  \beta H(\sigma,p) },
\label{ss-dis}
\end{equation}
where $Z_{\rm tot}(\beta)=\sum_{\sigma,p} e^{- \beta H(\sigma,p) }$.

\section{Equilibrium statistical mechanics} \label{SM}

In this section, for reference in studying dynamical
processes, we confirm some results of equilibrium
statistical mechanics using the canonical ensemble
\begin{equation}
P_{\rm can}(\sigma)= \frac{1}{Z(\beta)}
e^{- \frac{\beta}{2}\sum_{i\in\Lambda}\sum_{j \in B_i}\delta(\sigma_i,\sigma_j)}
\end{equation}
for the configuration space of $\sigma$, where $P_{\rm can}(\sigma)=\sum_p P_{\rm can}(\sigma,p)$.
We study
\begin{equation}
m= \lim_{|\Lambda| \to \infty}\sum_{\sigma} P_{\rm can}(\sigma)\frac{\sum_i  \delta(\sigma_i, 1) }{|\Lambda|}
\end{equation}
with an infinitely small external potential  $-\sum_i h_{\rm ex}\delta(\sigma_i,1)$ in the Hamiltonian,
and the free energy density defined as
\begin{equation}
f= -\lim_{|\Lambda| \to \infty}\frac{T}{|\Lambda|} \log Z.
\end{equation} 
It should be noted that the free energy density $f$ and the partition function
$Z$ are defined in the configuration space of $\sigma$.

In the calculation of $m$ and $f$, we conjecture that the
contribution from loops in the lattice can be ignored in
the large-size limit \cite{ORT,Mezard,Dembo}.
On the basis of this conjecture, the thermodynamic phase can be determined
using the model on a Cayley tree with three branches corresponding to coordination number 4. 
Concretely, it has been known that $m$ and $f$ are calculated
from the probability of the state $\sigma \in \{1,\cdots, q\}$
at a site connected with a cavity site, which is denoted as
$u(\sigma)$.
As shown in Appendix \ref{app-sec:SM}, we first have the self-consistent
equation for $u(\sigma)$,
\begin{align}
u(\sigma)=\frac{\left[\gamma u(\sigma)+1\right]^3}{\sum_\sigma
\left[\gamma u(\sigma)+1\right]^3 },
\label{q-sc}
\end{align}
with $\gamma=e^\beta-1$. Using the solutions of (\ref{q-sc}),
we express the order parameter $m(\beta)$ and the free energy density
$f(\beta)$ as
\begin{align}
m(\beta)&=\frac{[\gamma u(1)+1]^4}{\sum_\sigma [\gamma u(\sigma)+1]^4},
\label{m-rg} \\
f(\beta)&=-\beta^{-1} \log \frac{\sum_\sigma[\gamma u(\sigma)+1]^4}
{\left[\gamma \sum_\sigma u^2(\sigma)+1 \right]^2}.
\label{f-rg}
\end{align}

\begin{figure}[H]
    \subfigure[]{%
        \includegraphics[clip, width=0.45\columnwidth]{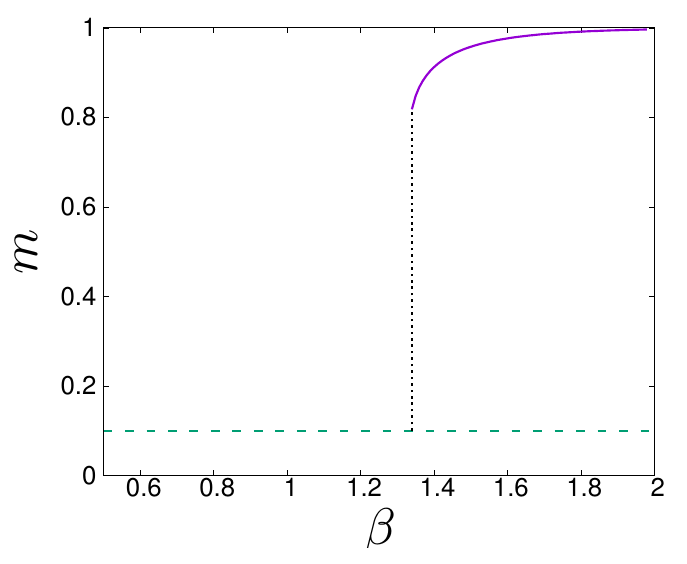}
        \label{mx_rand}	
         }%
    \subfigure[]{%
        \includegraphics[clip, width=0.5\columnwidth]{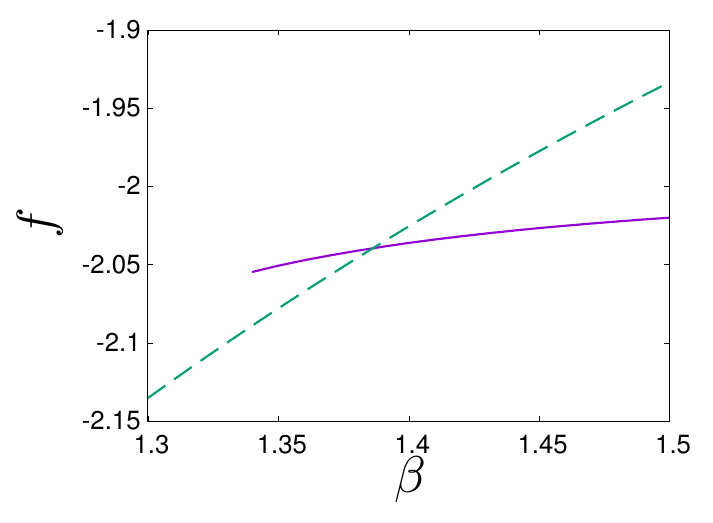}
      \label{freeenergy_rand}
        }%
    \caption{(a) $m$ as a function of $\beta$ for the case $q=10$. The solid (purple) and dashed (green) line represent $m_*(\beta)$ and $m_0(\beta)$, respectively. The vertical line indicates $\beta=\beta_{\rm sp}$. (b) $f$ as a function of $\beta$ for the case $q=10$. The solid (purple) and dashed (green) line represent $f_*(\beta)$ and $f_0(\beta)$, respectively. }
\end{figure}

Here we notice that (\ref{q-sc}) has the trivial solution
$u_0(\sigma)\equiv 1/q$ for any $\beta$. There exists
a temperature $\beta_{\rm sp}$ beyond which (\ref{q-sc})
has another solution denoted as $u_*(\sigma)$, where 
\begin{align}
u_*(\sigma)=
\begin{cases}
c_*(\beta) & (\sigma=1)  \\
\frac{1-c_*(\beta)}{q-1} & (2\leq \sigma \leq q)
\end{cases}
\end{align}
with $c_* \not = 1/q$. The temperature $\beta_{\rm sp}$ is called
the spinodal point. Using these two solutions $u_0$ and $u_*$,
we have $(m_0(\beta),m_*(\beta))$ and $(f_0(\beta),f_*(\beta))$.
We display  $(m_0(\beta),m_*(\beta))$ in  Fig.~\ref{mx_rand}, and 
$(f_0(\beta),f_*(\beta))$ in Fig.~\ref{freeenergy_rand} for the case $q=10$.
The equilibrium transition
temperature $\beta_c$ is identified as $f_0(\beta_c)=f_*(\beta_c)$.

\begin{figure}[H]
    \subfigure[]{%
        \includegraphics[clip, width=0.5\columnwidth]{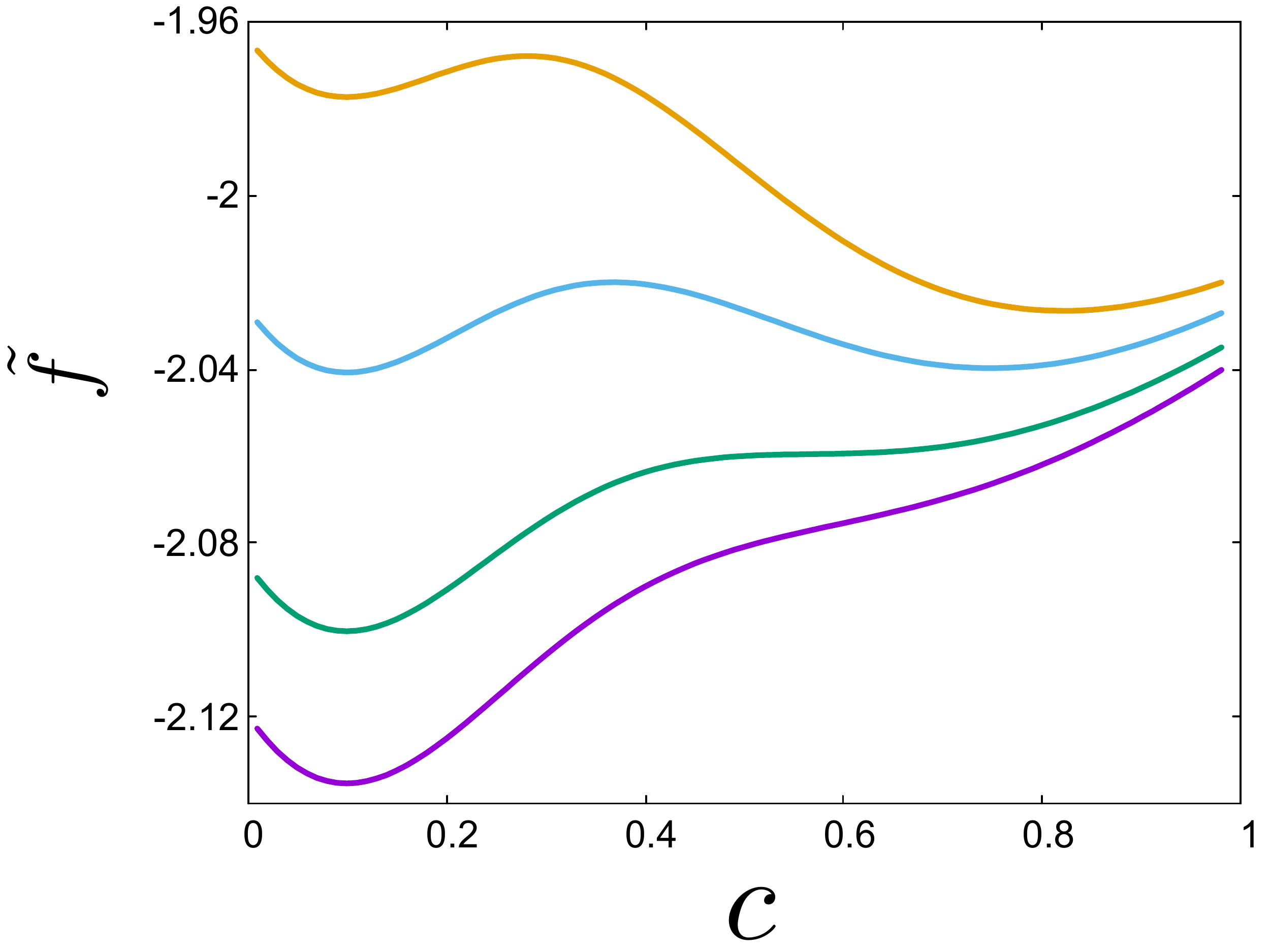}
         }%
    \subfigure[]{%
        \includegraphics[clip, width=0.5\columnwidth]{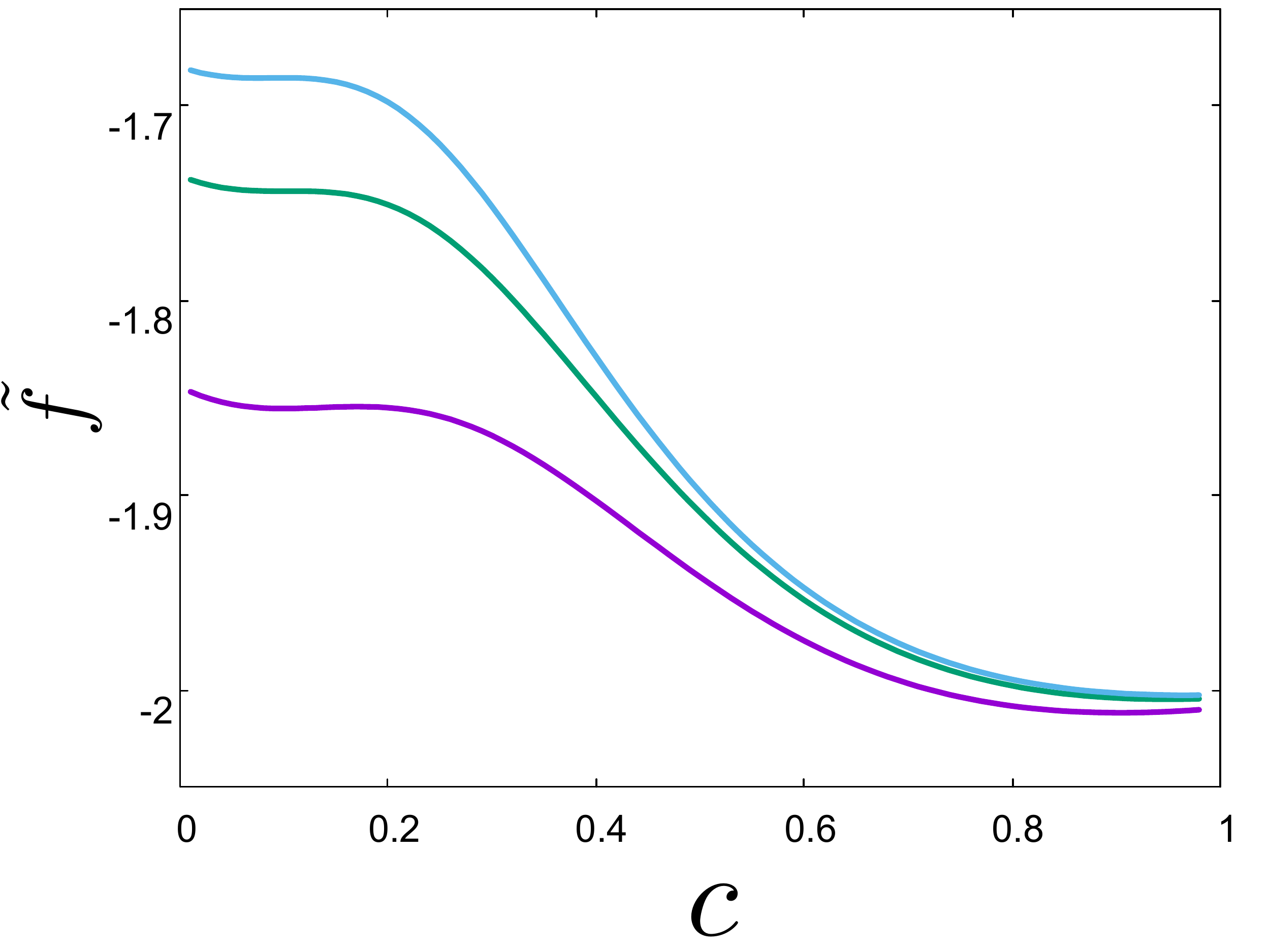}
        }%
   \caption{Free energy landscapes for (a) $\beta<\beta_{\rm sp}, \  \beta=\beta_{\rm sp}, \  \beta=\beta_{c}, \  $and $\beta_c<\beta<\beta_{\rm un}$ from below, and 
 (b) $\beta_c<\beta<\beta_{\rm un},  \  \beta=\beta_{\rm un},$ and $\    \beta>\beta_{\rm un}$ from below.}
 \label{lands_rand}    
 \end{figure}

To study the free energy landscape, we define
\begin{equation}
\tilde f(\beta,c)=
-\beta^{-1} \log
\frac{\sum_\sigma[\gamma \tilde u(\sigma)+1]^4}
{\left[\gamma \sum_\sigma \tilde u^2(\sigma)+1 \right]^2}, \label{tildef-def}
\end{equation}
where 
\begin{equation}
\tilde u(\sigma)=
\begin{cases}
c & (\sigma =1) \\
\frac{1-c}{q-1} & (2\leq \sigma \leq q).
\end{cases}
\end{equation}
Setting $c_0=1/q$, we have $f_0(\beta)= \tilde f(\beta,c_0)$
and $f_*(\beta)= \tilde f(\beta,c_*)$. Furthermore, by 
straightforward calculation, we confirm 
\begin{equation}
\frac{\partial}{\partial c}\left(e^{-\beta \tilde f(\beta,c)} \right) =0  \label{dp_dc_0}
\end{equation}
at $c=c_0$ and $c=c_*$.
See Appendix~\ref{app-sec:der-dp_dc_0} for the derivation.
Therefore, displaying $\tilde f(\beta,c)$ in Fig.~\ref{lands_rand},
we see that the solution $u_*$ appears at the spinodal point
$\beta=\beta_{\rm sp}$, and the equilibrium transition  occurs
at $\beta=\beta_c$. We also find that
$\partial^2 \tilde f/\partial c^2=0$ at $c=c_0$  and
$\beta=\beta_{\rm un}$, which means that the trivial solution
$u_0$ loses stability. Letting $F(u)$ be the right-hand side
of (\ref{q-sc}), the solutions of the self-consistent equation
are given by the fixed points $u_{n+1}=F(u_n)$. The temperature 
$\beta_{\rm un}$ is also characterized by the onset of the
instability of the trivial solution for this iteration
equation. For the case $q = 10$,
we find that $\beta_{\rm sp} \simeq 1.34$, $\beta_{c}\simeq 1.39$, and $\beta_{\rm un}\simeq 1.78$.

\section{Numerical simulation}\label{numerics}

\begin{figure}[H]
\centering
\includegraphics[width=0.47\columnwidth]{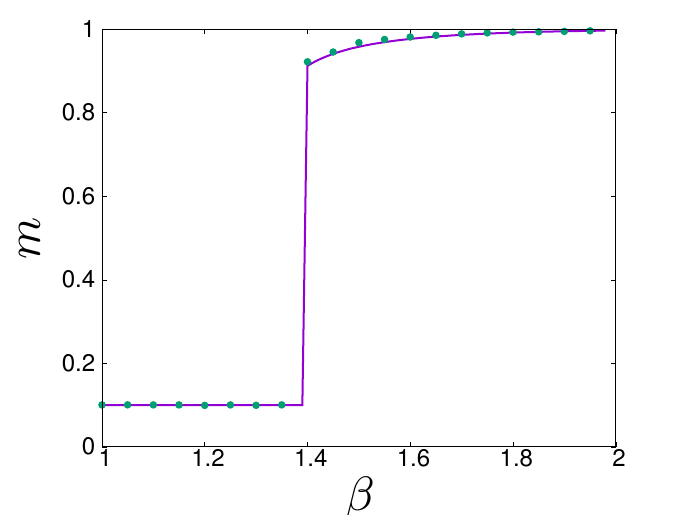}
\caption{Circle (green) symbols represent the simulation result
for the model given in Sec.~\ref{Model}, with $T_L=T_R=\beta^{-1}$.
The solid (purple) curve is the theoretical result obtained in Sec.~\ref{SM}.}
\label{rand_quench}
\end{figure}

Throughout this section, we simulate the model with $q=10$
and choose the numerical parameter $dp=0.01T_c$. 
Before presenting our results,
in Fig.~\ref{rand_quench}, we show the statistical average
of $\sum_i \delta(\sigma_i,1)/|\Lambda|$ for the model given
in Sec.~\ref{Model}, where $T_L=T_R=\beta^{-1}$ and $L_x=L_y=100$.
The numerical result agrees  with the theoretical calculation in
Sec.~\ref{SM}. 

Then, we prepare a metastable ordered phase of the
temperature $T_L$ in $1\leq i_x \le L_x/2$ and a stable disordered
phase of the temperature $T_R$ in $L_x/2+1 \leq i_x \leq L_x$. 
Here, we fix $T_L=1.01T_c$ so that the ordered phase is metastable, and
$T_R $ is assumed to be greater than $T_c$ so that the disordered phase
is stable. We prepare configurations at  $t=0$ via the following steps.
First, we remove the coupling between $ i_x=L_x/2$ and $i_x=L_x/2+1$
by setting $B_i^+ =\emptyset$ for $i_x=L_x/2$.
Second, we prepare a configuration where  $\sigma_i=1$ and $p_i=T_L $
for  any $i$ satisfying $i_x \le L_x/2$, while   each  $\sigma_i$  is randomly
chosen with  equal probability and $p_i=T_R$ for any $i$ satisfying
$i_x>L_x/2+1$.
Third, we make the system evolve up to 50 MCS.
The reached configuration is set to be an initial state at $t=0$
in the following analysis.
Note that $s$ MCS corresponds to the $t=sL_xL_y$ step in the procedure
described in Sec.~\ref{Model}.

\begin{figure}[H]
    \subfigure[]{%
        \includegraphics[clip, width=0.5\columnwidth]{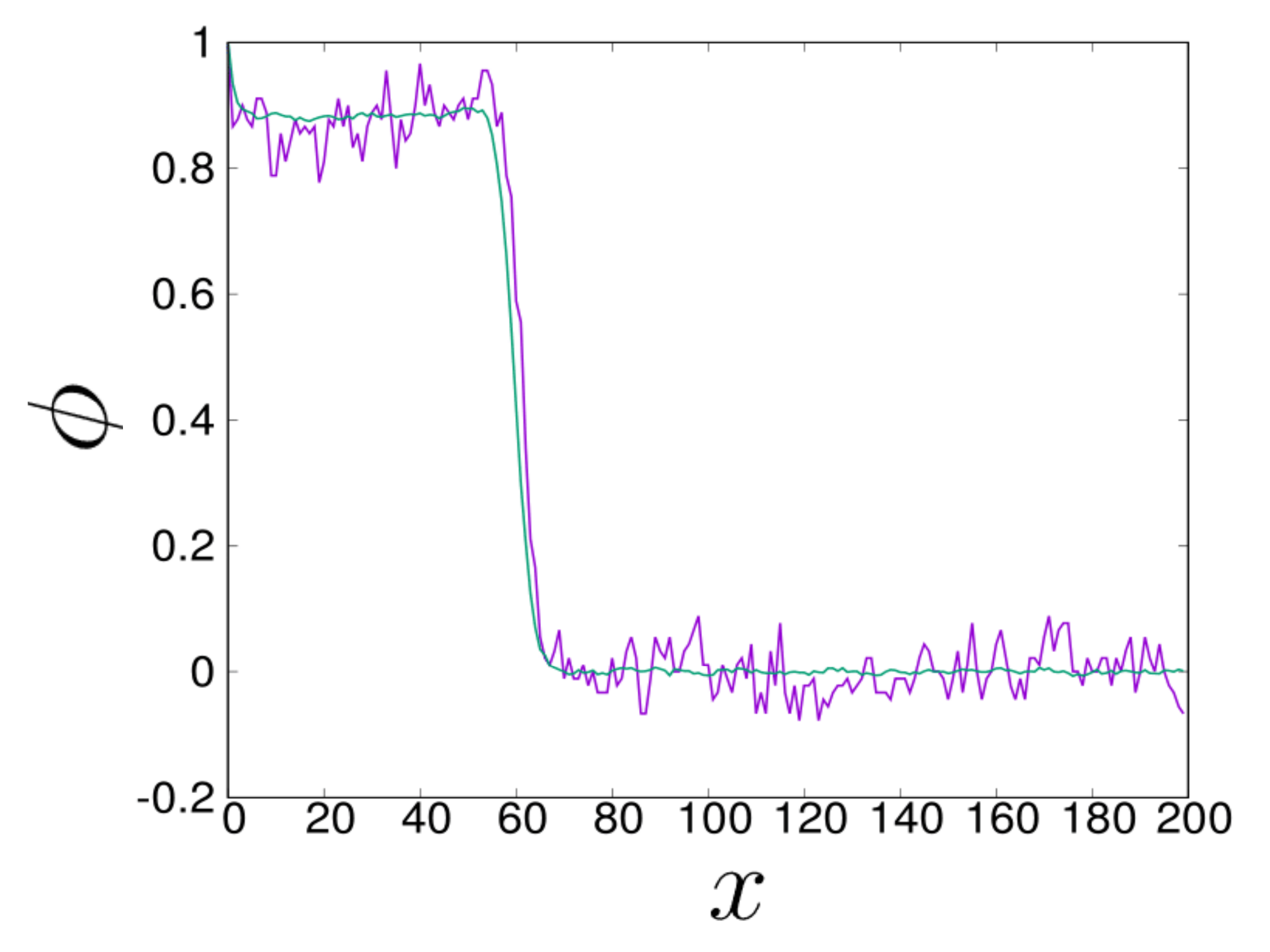}
        \label{phi_snapshot}
                 }%
    \subfigure[]{%
        \includegraphics[clip, width=0.5\columnwidth]{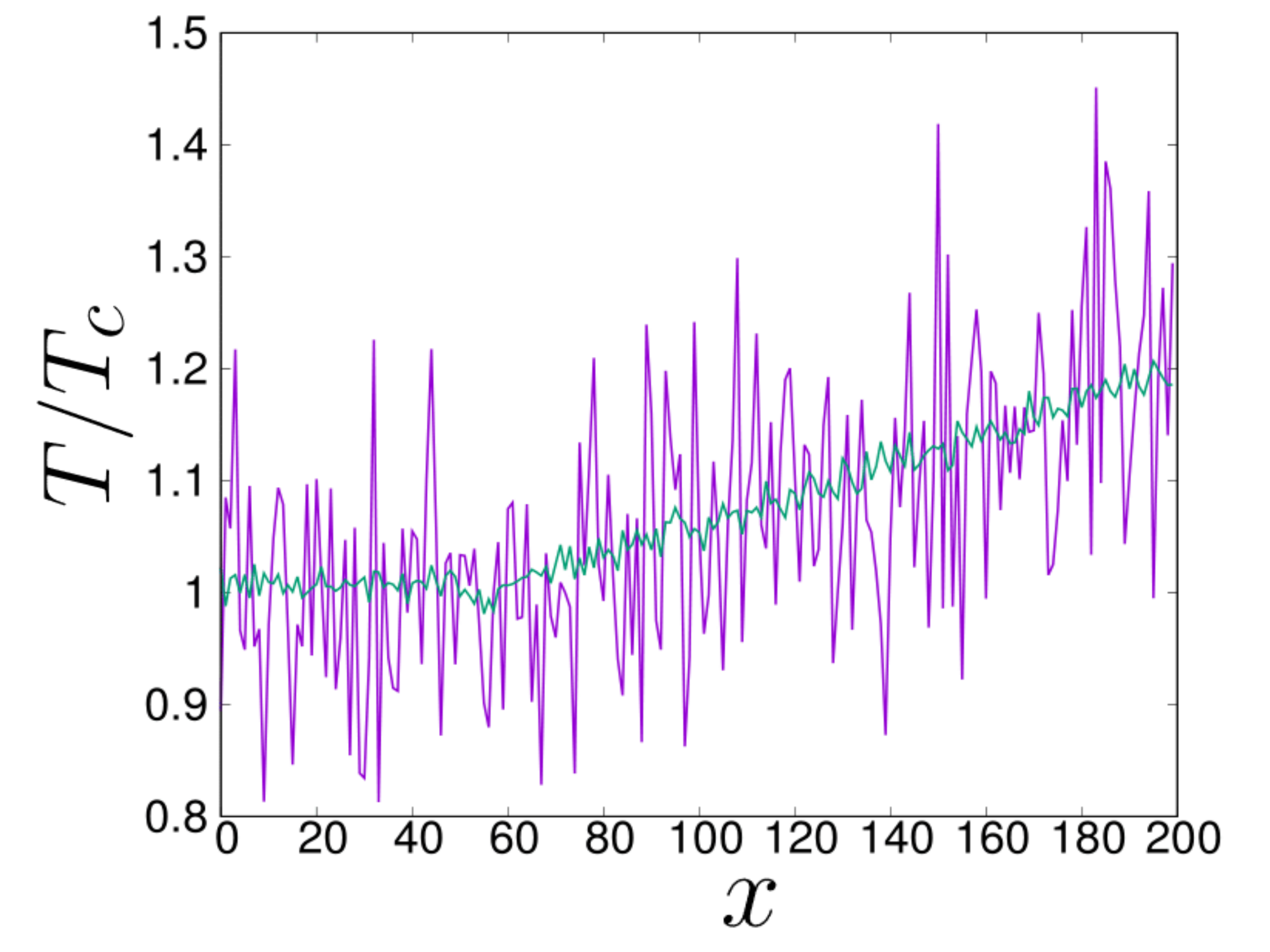}
      \label{tmp_snapshot}
        }%
  \caption{(a) $\phi(x,t)$ as a function of $x$ at $t=5\times 10^6$ MCS for the system with $L_x=200, L_y=100,T_{R}=1.2T_c$, and $T_{L}=1.01T_c$. The purple  and green graphs represent one snap shot and the average over 100 samples, respectively. (b) $T(x,t)/T_c$ as a function of $x$ for the same condition as (a). The colors of lines correspond to the graphs in (a).}
   \end{figure}

The stable disordered phase grows from the initial configuration. 
To describe the phase growth, we observe
an order parameter profile defined as
\begin{align}
\phi(x,t)\equiv \frac{q{m(x,t)-1}}{q-1}\label{phi}
\end{align}
with
\begin{align}
m(x,t) \equiv \frac{1}{L_y}\sum_{y=1}^{L_y} \delta(\sigma_{x,y},1)
\end{align}
for $1 \le x \le L_x$. We define the value $\phi(x,t)$ for any real $x$
via linear interpolation. 
Similarly, we define the temperature field as
\begin{align}
T(x,t)\equiv \frac{1}{L_y}\sum_{y=1}^{L_y}  p_{x,y} \label{T}.
\end{align}
Examples of $\phi(x,t)$ and $T(x,t)$ are displayed in Fig.~\ref{phi_snapshot} and ~\ref{tmp_snapshot}. 
 
\begin{figure}
\centering
\includegraphics[width=0.47\columnwidth]{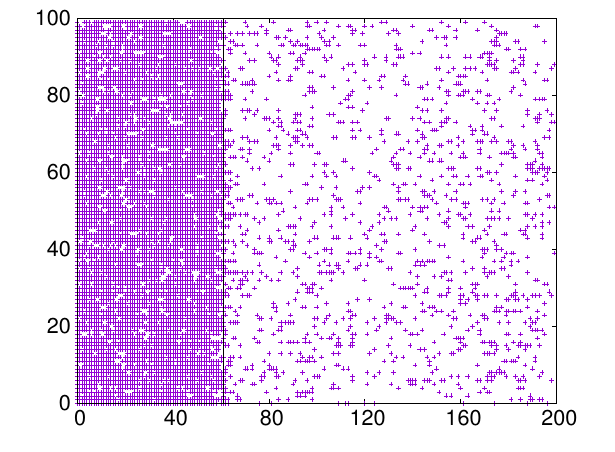}
\caption{ Configuration of $\sigma$ at $t=5\times10^6$ MSC for the system with $L_x=200, L_y=100$. The dots represent sites where $\sigma_i=1$.
The solid line represents the interface position.}
\label{gaikei}
 \end{figure}

 We identify the interface
position $X(t)$ from $\phi(x,t)$ such that $\phi(X(t),t)=0.5$. 
$X(t)$ is uniquely determined because $\phi$ crosses 0.5 only
once as far as we observe. $R(t)$ denotes the displacement of the interface $|X(t)-X(0)|$.
In Fig.~\ref{gaikei}, we show an example of configuration $\sigma$ with the interface position.

\begin{figure}[H]
\centering
\includegraphics[width=0.53\textwidth]{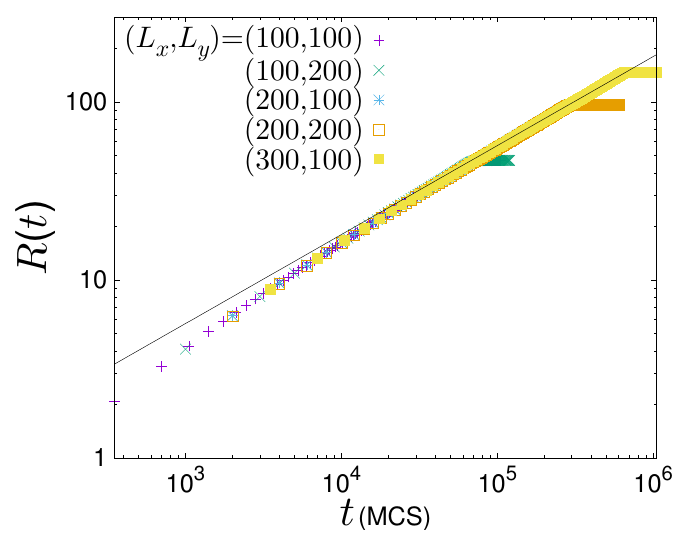}
\caption{$R(t)$ versus $t$ for $T_R=3T_c$.  The average of 100 samples was taken for the system with different system sizes.
The solid line  represents $R(t) = At^{0.5}$ with $A = 0.18$.}
\label{rand_result_Tr30}
\end{figure}

In Fig.~\ref{rand_result_Tr30}, we show the
statistical average of $R(t)$ for the case $T_R=3T_c>T_{\rm sp}$.
The five graphs correspond to results for different systems
sizes $(L_x,L_y)=(100,100)$, $(200,100)$, $(300,100)$, $(100,200)$,
and $(200,200)$.  Because the interface eventually reaches the left
boundary and does not move anymore, all the curves
become flat in the long-time limit. The guideline represents
$R(t) = At^{0.5}$ with $A = 0.18$. We also confirmed that
the $L_y$ dependence is hardly visible for this system size.
From these results, we reasonably
conjecture that the behavior in the large-$L_x$ limit is described
by the phase-field model because the extent of the metastability $\Delta$
is evaluated as $0.05$, as shown in Appendix~\ref{app-sec:latent heat}.  

\begin{figure}[H]
    \subfigure[]{%
        \includegraphics[clip, width=0.5\columnwidth]{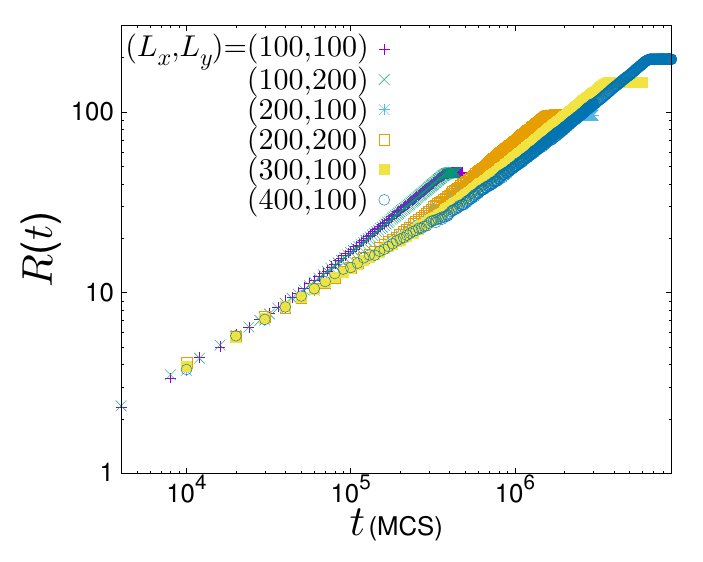}       
        \label{rand_result}
                 }%
    \subfigure[]{%
        \includegraphics[clip, width=0.5\columnwidth]{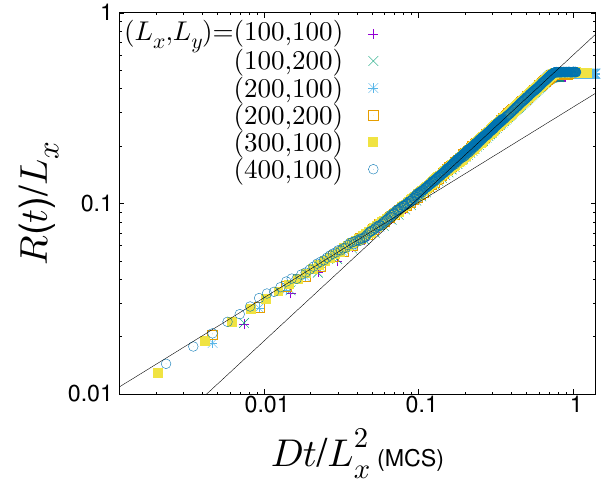}
      \label{rand_result_scaling}
        }%
\caption{ (a). $R(t)$ versus $t$ for $T_R=1.2T_c$. The average of 100 samples was taken for the system with different system sizes.  (b) $R(t)/L_x$ as a function of $z=Dt/L_x^2$. The two guidelines are $Az^{0.5}$ and $A'z^{0.75}$, where $A = 0.32$ and $A' = 0.60$,
respectively.}
 \end{figure}

An unexpected behavior of $R(t)$ is observed for the case $T_R=1.2T_c>T_{\rm sp}$. 
As shown in Fig.~\ref{rand_result}, four graphs for different system sizes,
$(L_x,L_y)=(100, 100), (200,100), (300,100)$, and $(400,100)$ do not overlap,
while the $L_y$ dependence is hardly visible. We then consider a finite size
scaling to plot $R/L_x$ as a function of $Dt/L_x^2$, where the
thermal diffusion constant $D$ is estimated as $1.9\times 10^{-2}$ from the measurement
of the relaxation time of the temperature field, as shown in
Appendix~\ref{app-sec:Diff}.
The result is displayed in Fig.~\ref{rand_result_scaling}. We find the scaling relation 
\begin{eqnarray}
\frac{R(t)}{L_x}=\bar{\mathcal{R}}\left(\frac{Dt}{L^2_x}\right)
\label{scaling}
\end{eqnarray}
works well, where $\bar{\mathcal{R}}$ is a scaling function. The
scaling function indicates the cross-over from $\bar{\mathcal{R}}(z)\simeq z^{0.5}$
to $\bar{\mathcal{R}}(z)\simeq z^{\alpha}$  with $0.5 \le \alpha \le 1$. It should be noted that the scaling relation in the late stage is not observed in the
phase-field model. 

\begin{figure}[H]
\centering
\includegraphics[width=0.53\textwidth]{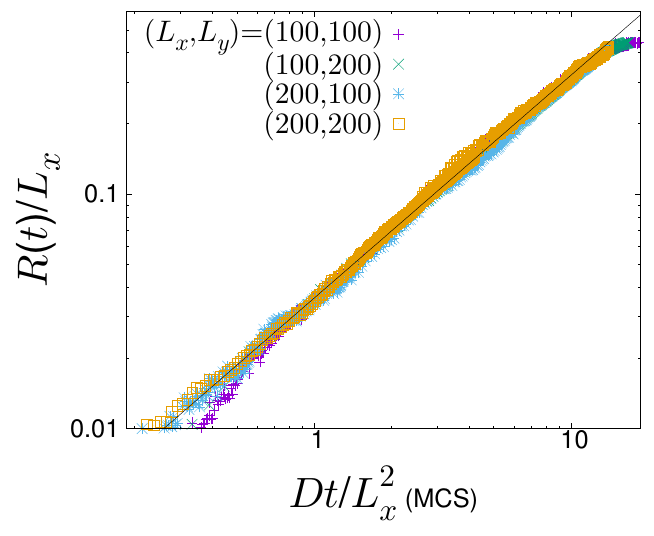}
\caption{$R /L_x$ as a function of $Dt/L_x^2$ for  the case $T_R=1.01T_c$.
The guideline represents $Az^{0.95}$ with $A = 3.6\times10^{-2}$.}
\label{rand_result_Tr101_scaling}
\end{figure}

The exponent $\alpha$ depends on the value of $T_R$.
Decreasing $T_R$ further, we find that the late stage becomes dominant
and $\alpha$ increases. For the case $T_R=T_L=1.01T_c<T_{\rm sp}$, $\alpha$ turns
out to be close to unity, as shown in Fig.~\ref{rand_result_Tr101_scaling}.

\begin{figure}[H]
    \subfigure[]{%
        \includegraphics[clip, width=0.33\columnwidth]{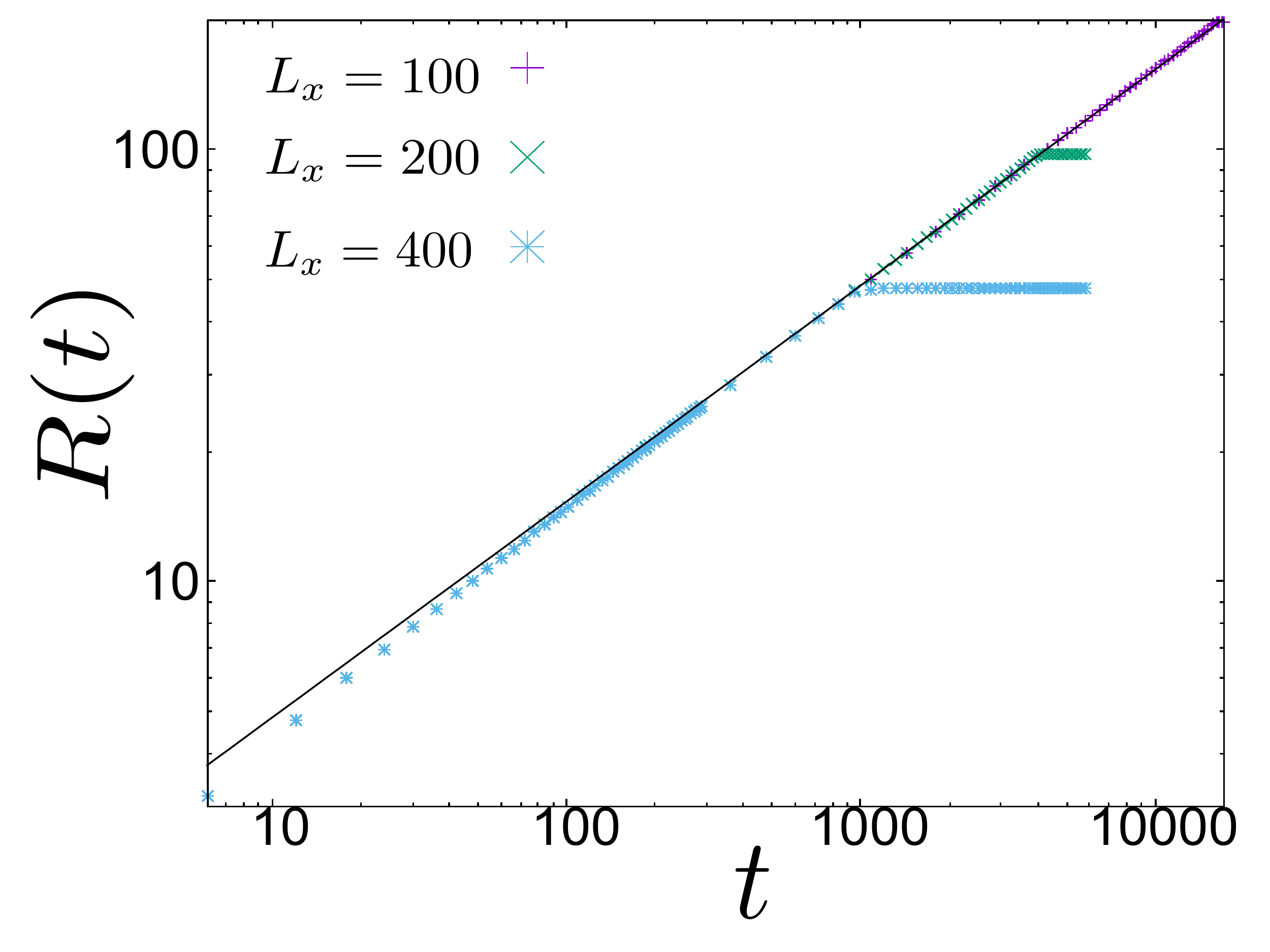}       
        \label{R_3}
                 }%
    \subfigure[]{%
        \includegraphics[clip, width=0.33\columnwidth]{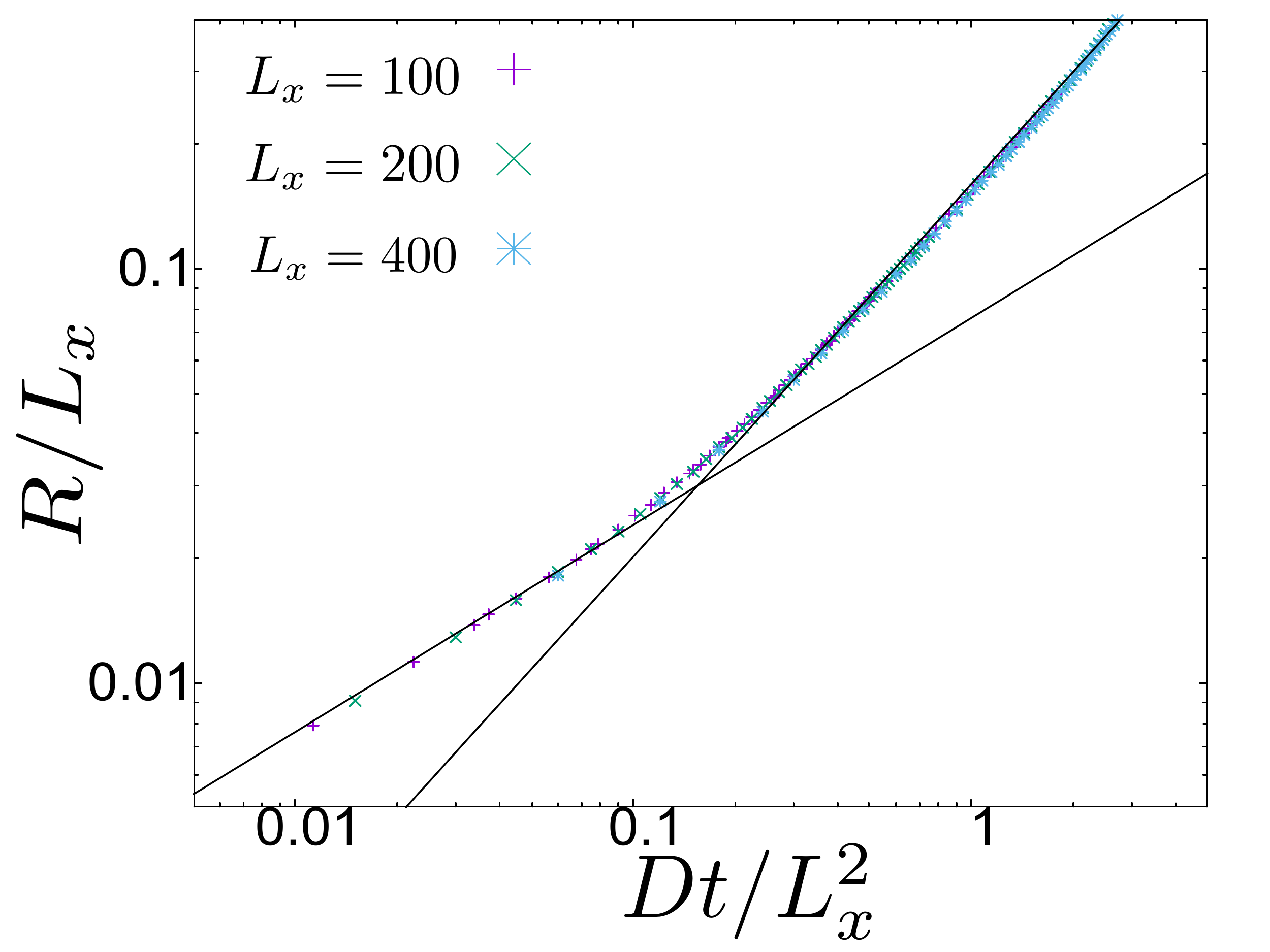}
      \label{R_005}
        }%
        \subfigure[]{%
        \includegraphics[clip, width=0.33\columnwidth]{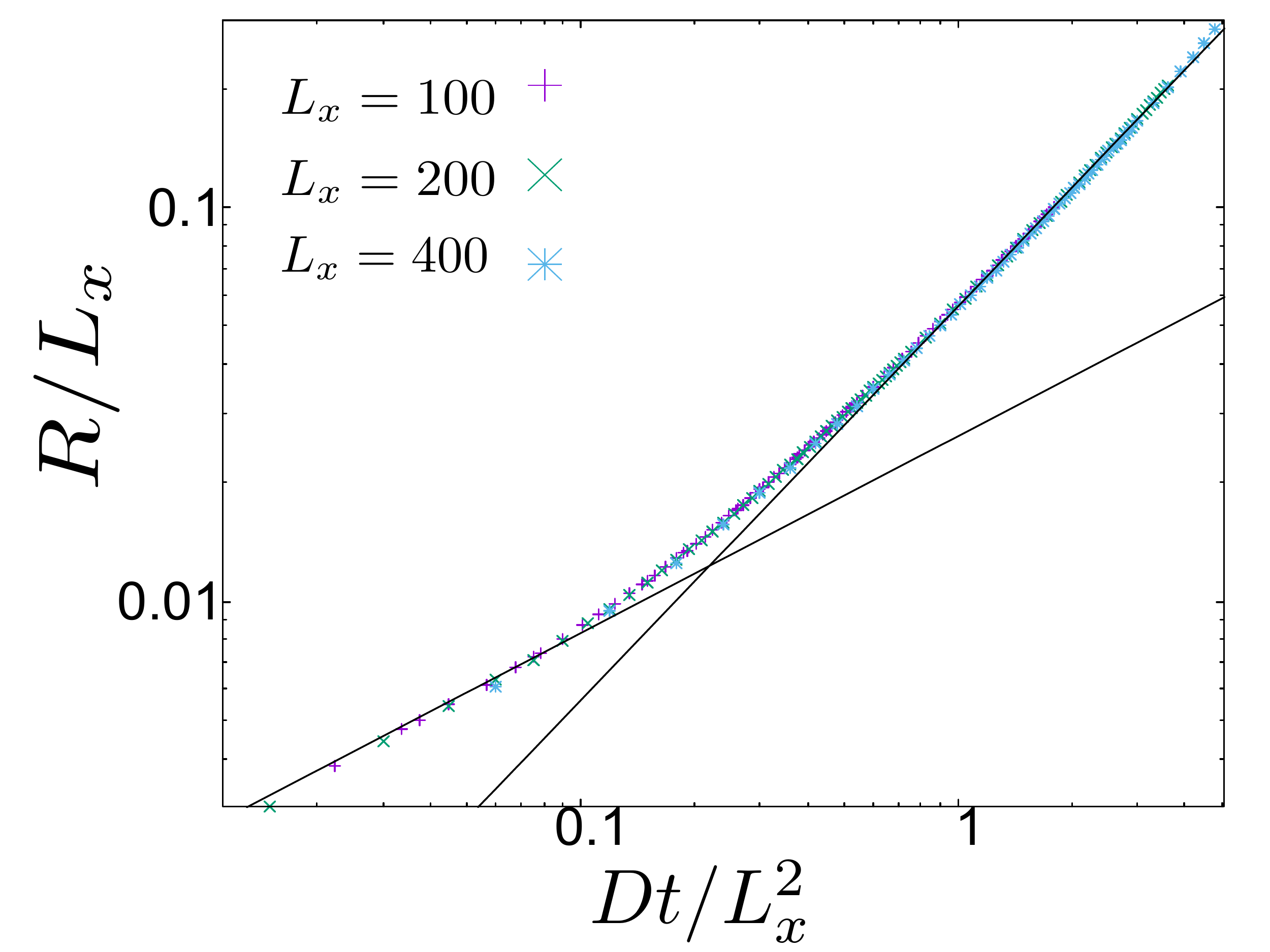}
      \label{R_001}
        }%
        \caption{Interface displacement for the phase-field model. (a) $R(t)$ as a function of $t$ for $T_R=4.0T_c$. The guideline represents $At^{0.5}$ with $A = 1.53$.
(b) $R/L_x$ as a function of $Dt/L_x^2$ for $T_R=1.05T_c$. 
The guidelines represent $Az^{0.5}$ and $A^\prime z^{0.9}$ with $A = 7.6\times 10^{-2}$ and $A^\prime =0.16$.
 (c) $R/L_x$ as a function of $Dt/L_x^2$ for $T_R=1.01T_c$.
 The guidelines represent $Az^{0.5}$ and $A^\prime z $ with $A = 2.6\times 10^{-2}$ and $A^\prime =  5.6\times 10^{-2}$.
 }
 \end{figure}

To our best knowledge, the cross-over behavior has
never been reported in previous studies. This observation raises a question about if this behavior is a unique feature of the stochastic model. In order to answer this question,
we carefully study finite-size effects for the deterministic
phase-field model. Specifically, we perform numerical simulations of
the model used in \cite{Karma-noise}.
 Concretely, we study the following equations of the order parameter field $\phi(x,t)$ and the dimensionless temperature field $\theta(x,t)$:
\begin{align}
\frac{\partial \phi}{\partial t}&=\partial_x^2 \phi +\phi-\phi^3 -\theta(1-\phi^2)^2,\\
\frac{\partial \theta}{\partial t}&= D\partial_x^2 \theta +\frac{1}{2}\frac{\partial p(\phi)}{\partial t},
\end{align}
where $\theta(x,t)$ is related to $T(x,t)$ as 
\begin{align}
\theta(x,t)= \frac{c_p}{T_c(\delta s)}{\left(T(x,t)-T_c\right)}.
\end{align}
The functional form of $p(\phi)$ is given by
\begin{align}
p(\phi)=\frac{15}{8}\left(\phi-\frac{2\phi^3}{3}+\frac{\phi^5}{5} \right).
\end{align}
We fix $T_L=1.01T_c$ and $D=1$.
The results are shown in Figs.~\ref{R_3}, \ref{R_005} and \ref{R_001}.
We observe the same behavior as that observed
in the statistical mechanics model. We thus conclude that 
the cross-over behavior is not specific to stochastic systems.

\begin{figure}[H]
\centering
\includegraphics[width=0.53\textwidth]{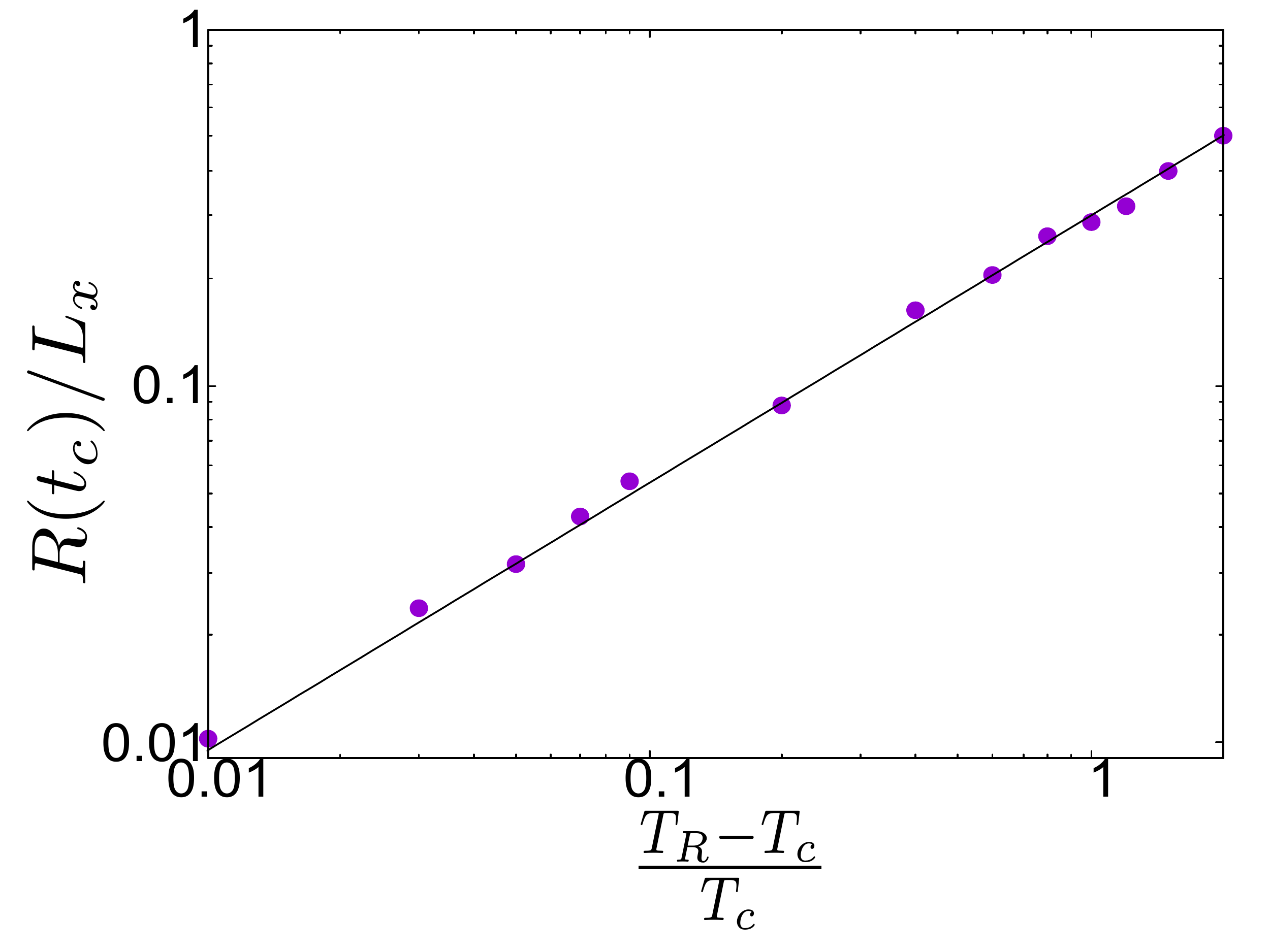}
\caption{$T_R$-dependence of $R(t_c)/L_x$. 
The guideline is given by (\ref{new-form}).}
\label{detail_analysis}
\end{figure}

Furthermore, for the phase-field model, we study $T_R$-dependence
more systematically. Let $t_c$ be the cross-over time. 
In Fig.~\ref{detail_analysis}, we plot $R(t_c)/L_x$  as a function of
$(T_R-T_c)/T_c$.  $R(t)$ is defined in the same way as done for the stochastic model. We then observe 
\begin{equation}
\frac{R(t_c)}{L_x}\simeq A \left(\frac{T_R-T_c}{T_c}\right)^{0.75}
\label{new-form}
\end{equation}
with $A=0.30$. Since the interface position moves up to $L_x/2$,
the interface reaches the boundary before the cross-over occurs
for temperatures $ T_R $ satisfying  ${R (t_c)} \geq L_x/{2}$.
In this case, only $R(t)\sim t^{0.5}$ is observed.
On the other hand, as the temperature $T_R$ is close to $T_c$ with
large $L_x$ fixed, $R(t_c)/L_x$ tends to zero. This means
that only $R(t) \sim t^\alpha$ with $\alpha \simeq 1$ is observed.
Note that if we take the limit 
$L_x \to \infty$ first, $R(t)\sim t^{0.5}$ is observed. However, when we observe the interface
motion in finite-size systems, the behavior depends on the order of
the three limits $L_x \to \infty$, $t \to \infty$, and $T_R \to T_c$.

\section{Concluding remarks}\label{remark}
We have proposed a $q$-state Potts model with an additional variable representing the kinetic energy at each site. 
By designing a two-dimensional lattice, where each site is sparse-randomly
  connected in one direction and local in the other direction,
we have explicitly calculated the thermodynamic properties via equilibrium statistical mechanics. Simulating this model, we have numerically observed that the interface between the stable and metastable phases moves following the scaling relation (\ref{scaling}) with the scaling function $\bar{\mathcal{R}}(z)$. The scaling function shows a cross-over from $\bar{\mathcal{R}}(z)\simeq z^{0.5}$ to $\bar{\mathcal{R}}(z) \simeq z^{\alpha}$, where the cross-over value of $z$ and the exponent $\alpha$ depends on the temperature of the heat bath attached to the stable phase.
We have only analyzed the case $q = 10$. Changing the value of
  $q$ influences the quantitative behavior through the $\Delta$-dependence.
  We conjecture that the $\Delta$ dependence in the large system size limit
  should be consistent with the theoretical result for the phase-field model
  \cite{HS}, because our stochastic model belongs to the same universality
  class as the phase-field model. 
In ending this paper, we make two remarks.

\begin{figure}[H]
    \subfigure[]{%
        \includegraphics[clip, width=0.31\columnwidth]{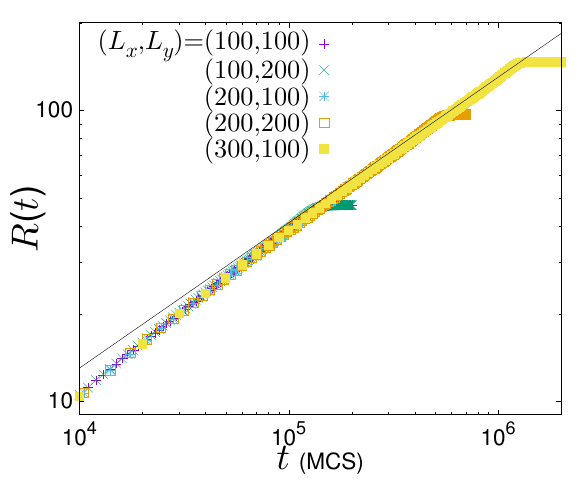}       
        \label{result_2d_Tr3}
                 }%
    \subfigure[]{%
        \includegraphics[clip, width=0.34\columnwidth]{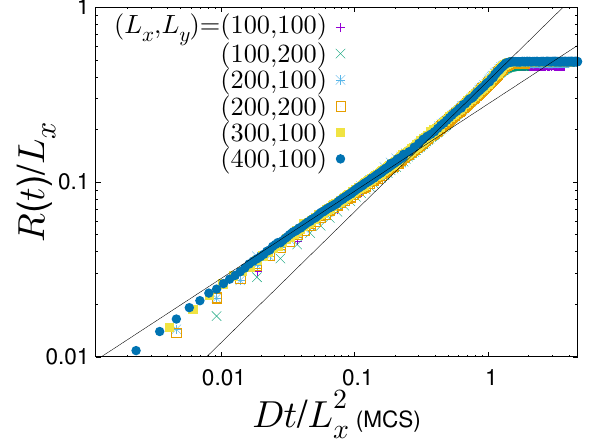}
      \label{result_2d_scaling}
        }%
        \subfigure[]{%
        \includegraphics[clip, width=0.33\columnwidth]{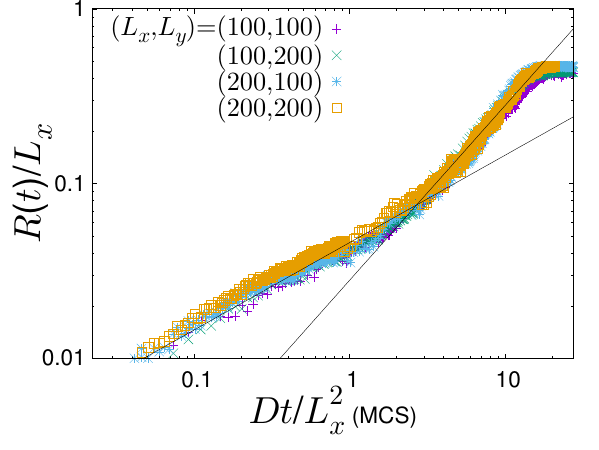}
      \label{result_2d_Tr1.01}
        }%
        \caption{Interface displacement of the $q$-state Potts model on the square lattice. The value of $D$ is measured  as $0.02$ in the same way as shown in Appendix \ref{app-sec:Diff}. The value of $T_c$ for $q=10$ was  calculated exactly \cite{Wu}.  (a) $R(t)$ as a function of $t$ for $T_R=3.0T_c$.
The guideline represents $At^{0.5}$ with $A=0.13$. 
(b) $R/L_x$ as a function of $Dt/L_x^2$ for $T_R=1.2T_c$.
The guidelines represent $Az^{0.5}$ and $A^\prime z^{0.75}$ with $A= 0.28$ and $A^\prime = 0.38$.
(c)$R/L_x$ as a function of $Dt/L_x^2$ for $T_R=1.01T_c$.  
The guidelines represent $Az^{0.5}$ and $A^\prime z $ with $A = 4.6\times 10^{-2}$ and $A^\prime =2.8\times 10^{-2}$.
 }
 \end{figure}

The first is on the status of our model. We note that  
our model is rarely realizable in experiments, like a more familiar
mean-field type model, where a site $(i_x,i_y)$ is connected to all sites
in the $i_x+1$ layer. Despite apparent unphysical nature
of the random lattice, the phase growth in our model is qualitatively
same as that for the model on the square lattice,
as shown in Figs.~\ref{result_2d_Tr3}, \ref{result_2d_scaling},
and \ref{result_2d_Tr1.01}.
This is a special property of our model, because the mean-field type
model shows a different behavior \cite{HOS2}. Thus, toward the microscopic
derivation of a coupled equation for the order parameter field and the
temperature field, numerical results for our model should be theoretically
explained by extending  the analysis shown in Sec.~\ref{SM} and
Appendix \ref{app-sec:SM}.

The second remark is on the Mullins-Sekerka instability \cite{MS}.
The instability of propagating interfaces was studied in the phase-field
model for the case $\Delta\geq 1$ \cite{Kupf,Braun}. We do not
clearly understand the instability condition for the case $\Delta <1$.
Nevertheless, in any cases, it is clear that the instability
  is  not observed 
in our model on the sparse-randomly connected lattice, because there is no
spatial correlation in the vertical direction. While this aspect
may be a disadvantage of the model, we note that the instability is not
our main concern. For the model
on the square lattice, there remains a possibility that Mullins-Sekerka
instability could appear for larger system sizes than we studied. Clarifying
the instability condition is left for future study.

\begin{acknowledgements}
This work was supported by KAKENHI (Grant Nos. 17H01148, 19H05795, and 20K20425).
\end{acknowledgements}

\begin{spacing}{0.5}
\noindent{\small {\bf Data Availibility} \ The datasets generated during and/or analysed during the current study are available from the corresponding author on reasonable request.}\\
\end{spacing}

\vspace{7pt}

\begin{spacing}{0.8}
\noindent{\small {\bf Conflict of Interest} \ \  The authors have no financial or proprietary interests in any material discussed in this article.
}
\end{spacing}

\appendix\normalsize
\renewcommand{\theequation}{\Alph{section}.\arabic{equation}}
\setcounter{equation}{0}
\makeatletter
  \def\@seccntformat#1{%
    \@nameuse{@seccnt@prefix@#1}%
    \@nameuse{the#1}%
    \@nameuse{@seccnt@postfix@#1}%
    \@nameuse{@seccnt@afterskip@#1}}
  \def\@seccnt@prefix@section{Appendix }
  \def\@seccnt@postfix@section{:}
  \def\@seccnt@afterskip@section{\ }
  \def\@seccnt@afterskip@subsection{\ }
\makeatother

\section{Derivation of the formulas in Sec. \ref{SM}}
\label{app-sec:SM}

In this section, we derive formulas 
(\ref{q-sc}), (\ref{m-rg}), (\ref{f-rg}),
and (\ref{dp_dc_0}) in Sec.~\ref{SM}.

\subsection{Derivation of (\ref{q-sc})}
\label{app-sec:der-q-sc}

\begin{figure}[H]
\centering
\includegraphics[width=0.35\textwidth]{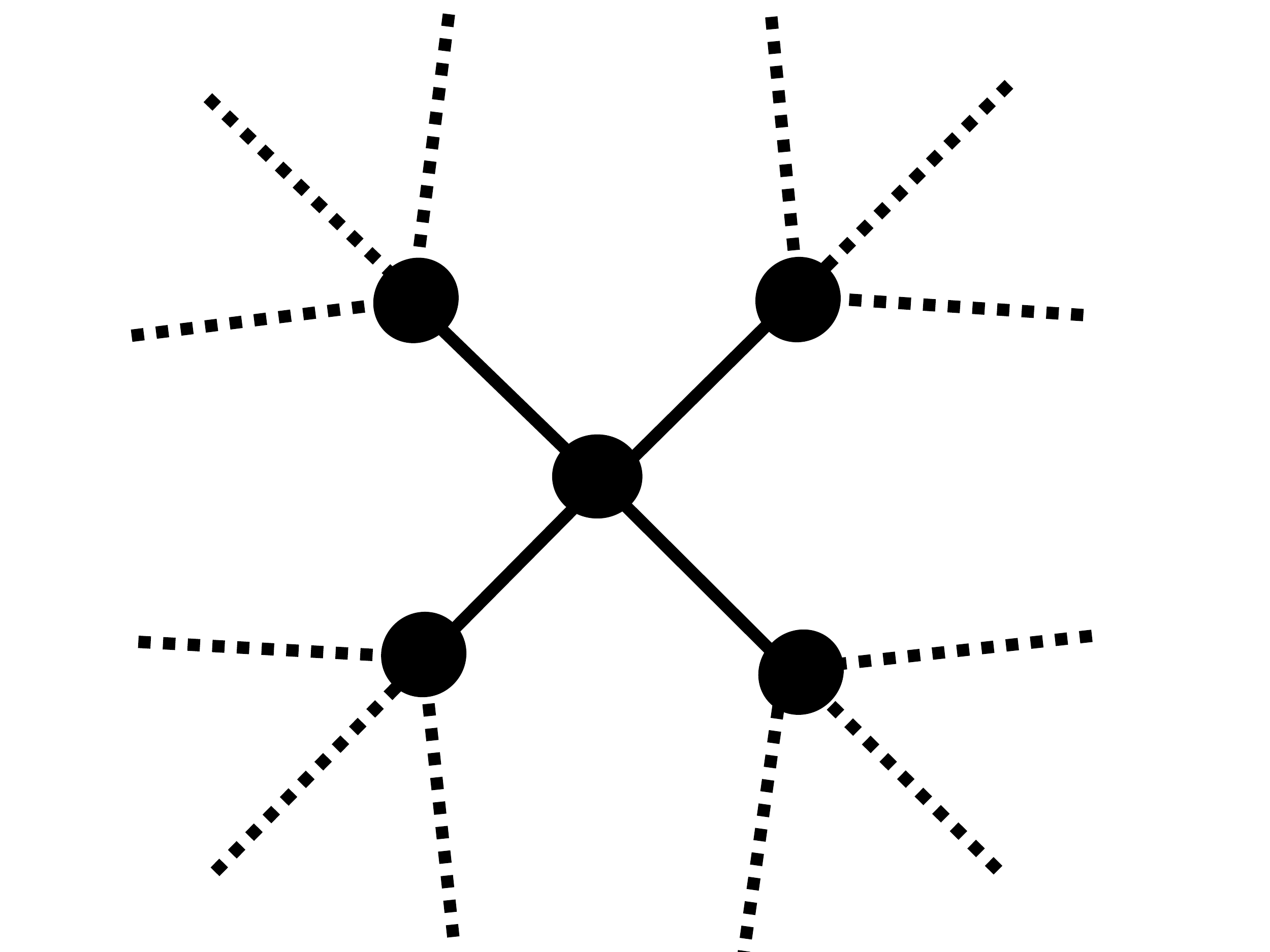}
\caption{Illustration of the Cayley tree.}
\label{app_caley}
\end{figure}

We study a Cayley tree with a root site connected with four
sites in the first generation. Each site in the $n$-th generation
$(n \ge 1)$  is  connected with three sites in the $n+1$-th generation.
See  Fig.~\ref{app_caley}  for the illustration of the Cayley tree. 

\begin{figure}[H]
\centering
\includegraphics[width=0.35\textwidth]{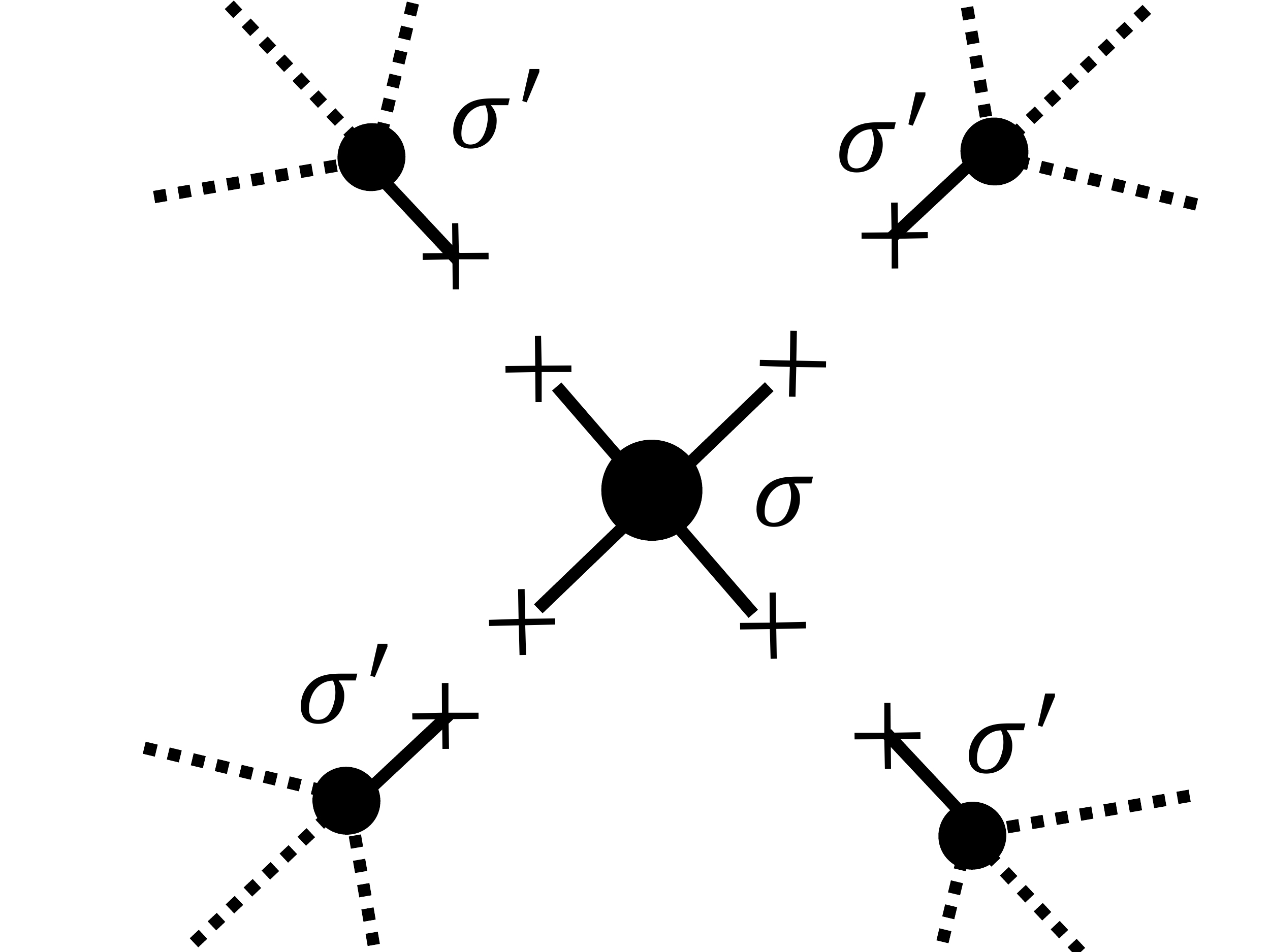}
\caption{Graphical representation of (\ref{Z}).}
\label{app_caley1}
\end{figure}

Let $Z$ be
the partition function of the Potts model on the lattice. We 
consider the partition function of a system in which 
a root site is replaced by the cavity and the state of a site
in the first generation is fixed as $\sigma' \in \{1,\cdots, q\}$,
which is denoted by $\tilde Z_1(\sigma')$. $Z$ is then the partition function of the model expressed as
\begin{equation}
Z= \sum_{\sigma} \left(\sum_{\sigma'}
e^{\beta\delta(\sigma,\sigma')} \tilde Z_1(\sigma')\right)^4.
\label{Z}
\end{equation}
A graphical representation is displayed in Fig.~\ref{app_caley1}. 

\begin{figure}[H]
\centering
\includegraphics[width=0.4\textwidth]{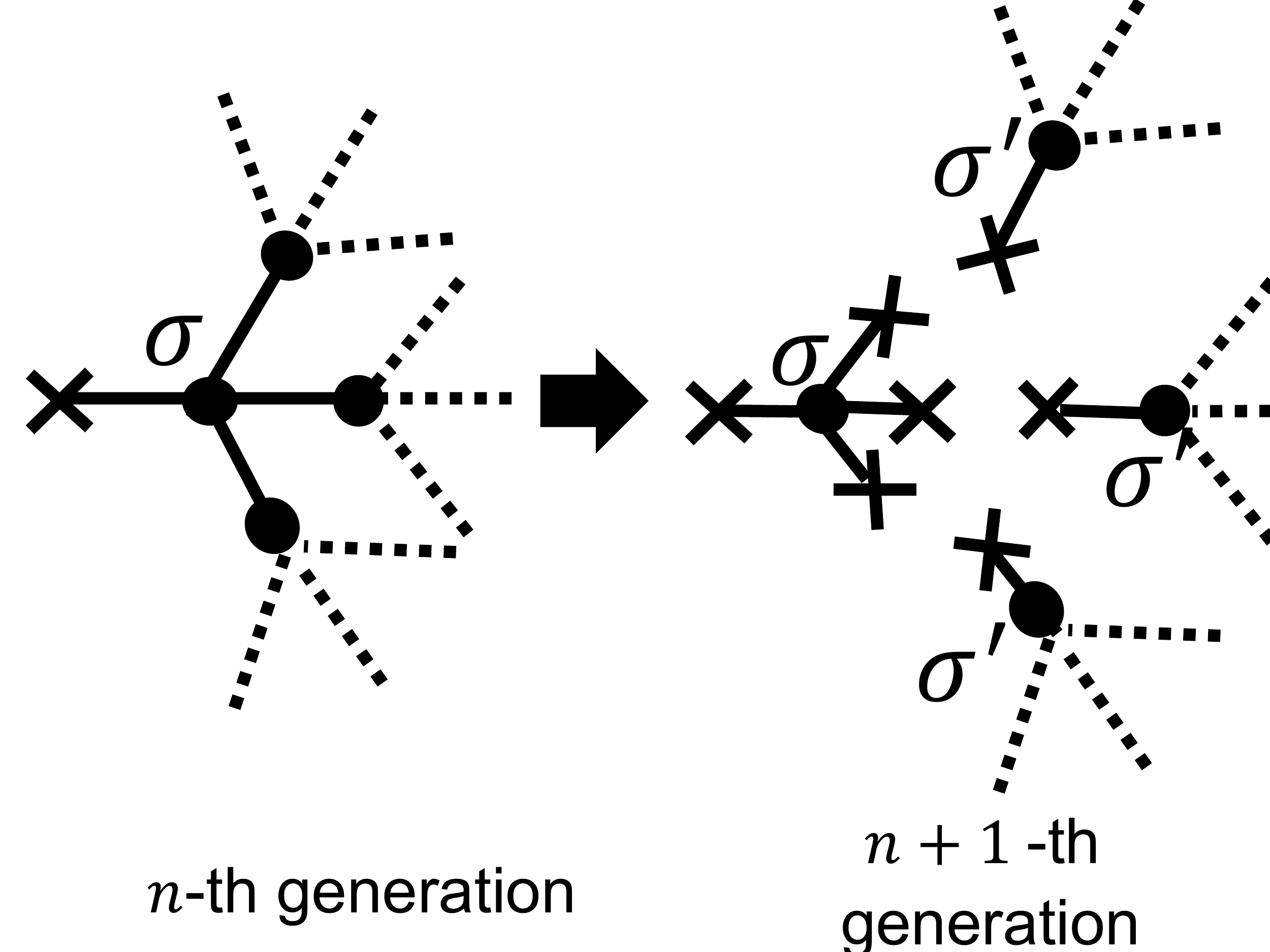}
\caption{Graphical representation of (\ref{Z-itr}).}
\label{app_rand_0}
\end{figure}

By setting
\begin{align}
\gamma&\equiv e^\beta-1,\\
G_1 &\equiv \sum_\sigma \tilde{Z_1}(\sigma) ,
\end{align}
we rewrite (\ref{Z}) as
\begin{equation}
Z= \sum_{\sigma} \left( \gamma \tilde Z_1(\sigma)+G_1 \right)^4.
\label{Z-2}
\end{equation}
Defining $\tilde Z_n(\sigma)$ and $G_n$ similarly,
we have the iterative equation
\begin{equation}
\tilde Z_{n}(\sigma)= \left(\gamma \tilde Z_{n+1}(\sigma)+G_n \right)^3,
\label{Z-itr}
\end{equation}
whose graphical representation is shown in Fig.~\ref{app_rand_0}.

We define $u_n(\sigma)$ as
\begin{align}
u_n(\sigma)\equiv \frac{\tilde{Z_n}(\sigma)}{G_n}, \label{defu}
\end{align}
which corresponds to the probability of the state $\sigma$ of
the cavity-connected site in the $n$-th generation. 
By substituting (\ref{defu}) into (\ref{Z-itr}), we obtain
\begin{align}
G_n u_n(\sigma)&=G_{n+1}^3 \left[\gamma u_{n+1}(\sigma) +1\right]^3.
\label{us}
\end{align}
We also have 
\begin{align}
G_n=G_{n+1}^3\sum_\sigma
\left[\gamma u_{n+1}(\sigma)+ 1\right]^3
\label{us_normalize}
\end{align}
using $\sum_{\sigma} u_n(\sigma)=1$.
From (\ref{us}) and (\ref{us_normalize}),
we derive the iterative equation for $u_n(\sigma)$,
\begin{align}
u_n(\sigma)=\frac{\left[\gamma u_{n+1}(\sigma)+1\right]^3}
{\sum_\sigma \left[\gamma u_{n+1}(\sigma)+1\right]^3 }.
\label{u}
\end{align}
Assuming homogeneity in the equilibrium state,
$u_n(\sigma)$ is independent of $n$ in the large-size limit.
This provides (\ref{q-sc}).

\subsection{Derivation of (\ref{m-rg})} 
\label{app-sec:der-m-rg}

The order parameter $m$ for the model is calculated by
the expectation value of $\delta(\sigma,1)$ at the root site.
That is, 
\begin{align}
m=\frac{1}{Z}\sum_{\sigma}\delta(\sigma,1)
\left[\gamma \tilde Z_1(\sigma) + G_1 \right]^4.
\end{align}
Using the expression given in (\ref{Z-2}), we have
\begin{align}
m=\frac{[\gamma u_1(1)+1]^4}{\sum_\sigma [\gamma u_1(\sigma)+1]^4}.
\end{align}
By replacing $u_1$ with the solution of the self-consistent equation
(\ref{q-sc}), we obtain (\ref{m-rg}).

\subsection{Derivation of (\ref{f-rg})}
\label{app-sec:der-f-rg}

\begin{figure}[H]
\centering
\includegraphics[width=0.4\textwidth]{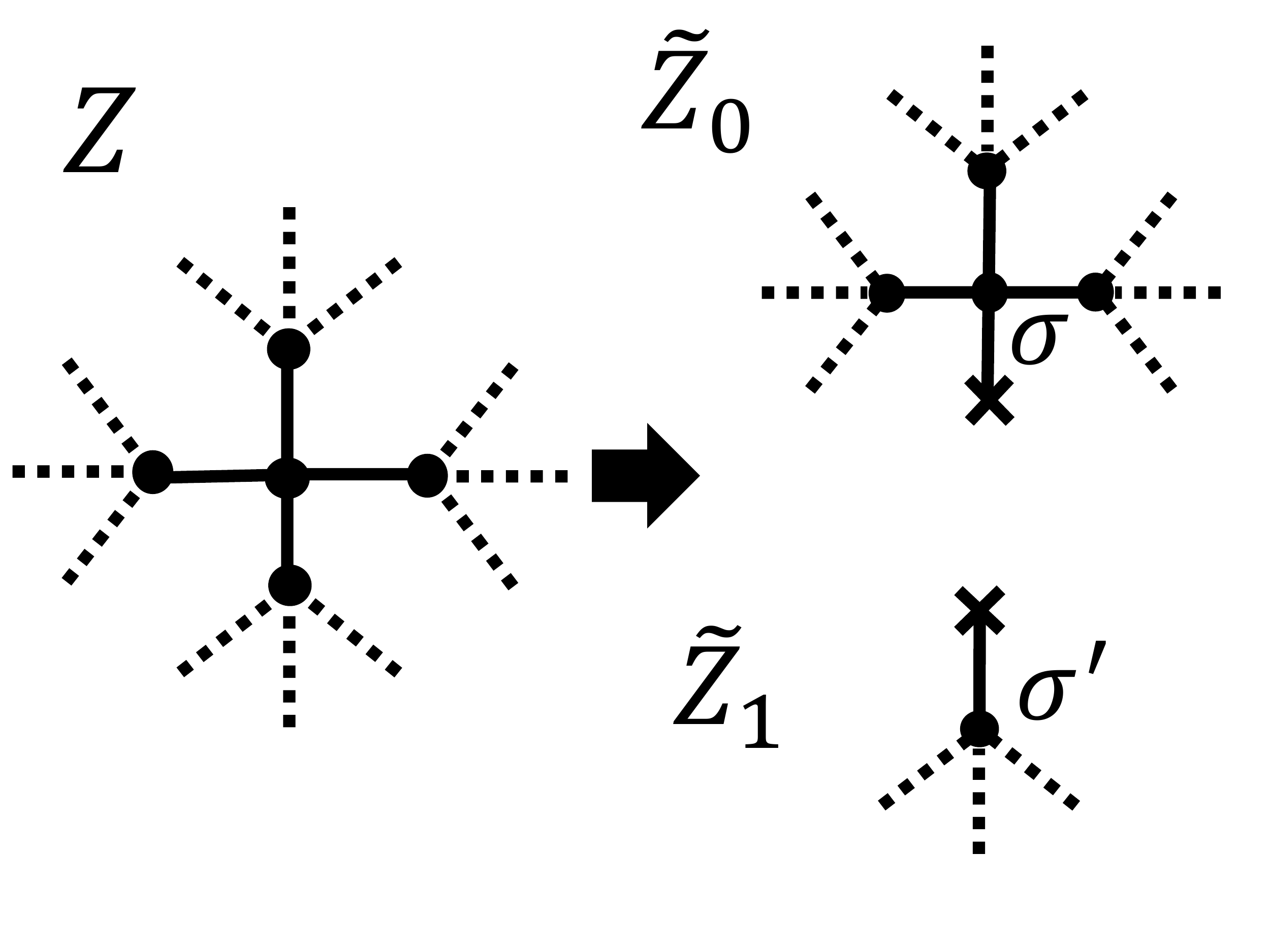}
\caption{By removing one edge, we get two rooted graphs. $\times$ represents the cavity.}
\label{app_rand_1}
\end{figure}

To derive the free energy density, we use a tactical
method manipulating graphs. We first remove one edge
connected to the root site. The partition function
of this system with $\sigma$ at the root site and
$\sigma'$ at the other site connected by the removed
edge is $\tilde{Z}_0(\sigma)\tilde{Z}_1(\sigma')$.
See Fig.~\ref{app_rand_1}. We thus express the partition function $Z$ as
\begin{align}
Z&=\sum_{\sigma,\sigma^\prime} e^{\beta \delta(\sigma,\sigma^\prime)}
\tilde{Z_0}(\sigma)\tilde{Z_1}(\sigma^\prime)\\
&=G_0G_1\left[\gamma \sum_\sigma u_0(\sigma)u_1(\sigma)
+1 \right]
\label{Z0},
\end{align}
where $G_0\equiv \sum_{\sigma}\tilde{Z}_0(\sigma)$ and
$u_0(\sigma)\equiv \tilde{Z}_0(\sigma)/G_0$.
Note that $u_0(\sigma)$ also satisfies (\ref{u}).

\begin{figure}[H]
\centering
\includegraphics[width=0.4\textwidth]{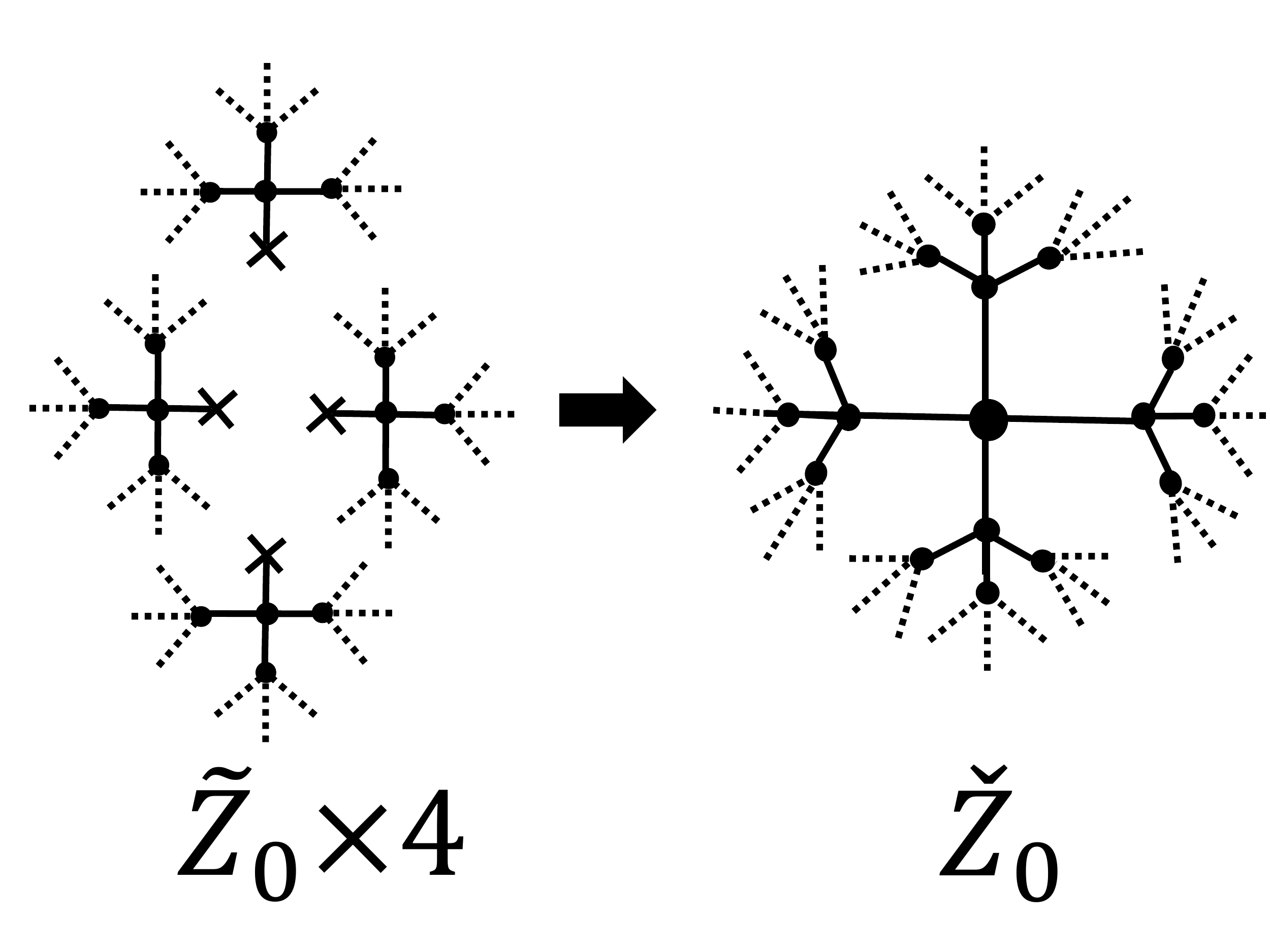}
\caption{By adding one site, we combine four graphs with the root site.}
\label{app_rand_2}
\end{figure}

Next, we prepare four independent systems. The partition
function of the total system is $Z^4$. We remove one
edge connected to the root site for each graph. Then,
we combine four graphs with the root site by adding
one site. See Fig.~\ref{app_rand_2}. The partition function of this system, $\check{Z_0}$,
is expressed as 
\begin{align}
\check{Z_0}=G_0^4\sum_\sigma\left[\gamma u_0(\sigma)+1
\right]^4
\label{checkZ0}.
\end{align}

Similarly, another Cayley tree is obtained by combining
the other graphs with another added site, and the partition
function is written as 
\begin{align}
\check{Z_1}=G_1^4\sum_\sigma\left[\gamma u_1(\sigma)+1
\right]^4
\label{checkZ1}.
\end{align} 
The free energies of the original system and the new system
are $-T\log Z^4$ and $-T\log \check{Z_0}\check{Z_1}$, respectively.
The difference in free energy is equal to $2f$,
where $f$ is the free energy density,
because the two systems have the same free energy density in the large-size limit
and the new system is the original system with two sites added.
That is,
\begin{align}
- T\log \check{Z_0}\check{Z_1} + T\log Z^4 = 2f. 
\end{align}
This is rewritten as 
\begin{align}
&e^{-\beta f}=\left(\frac{\check Z_0\check Z_1}{Z^4}\right)^{1/2}\\
&=\left(\frac{\sum_{\sigma'}[\gamma u_0(\sigma')+1]^4
         \sum_{\sigma''}[\gamma u_1(\sigma'')+1]^4}
{\left[\gamma \sum_\sigma u_0(\sigma)u_1(\sigma)+1 \right]^4}
\right)^{1/2}.
\end{align}
By replacing $u_0$ and $u_1$ with the solution of the self-consistent
equation (\ref{q-sc}), we obtain (\ref{f-rg}).

\subsection{Derivation of (\ref{dp_dc_0})}
\label{app-sec:der-dp_dc_0}

For later convenience, we set
\begin{equation}
\tilde c \equiv \frac{1-c}{q-1}.
\label{tilde-c}
\end{equation}
From the definition of $\tilde f$ given in (\ref{tildef-def}),
the left-hand side of (\ref{dp_dc_0}) is calculated as 
\begin{align}
&\frac{\partial}{\partial c} e^{-\beta \tilde f(\beta,c)} \nonumber \\
&=4\gamma \frac{ (\gamma c+1)^3-(\gamma \tilde c+1)^3}{(\gamma c^2+\gamma(q-1)\tilde c^2+1)^2} \nonumber\\
&-4\gamma (c-\tilde c )  \frac{(\gamma c+1)^4+ (q-1)(\gamma\tilde c+1)^4 }{ (\gamma c^2+\gamma(q-1)\tilde c^2+1)^3 }.
\label{app:dp_dc_0_1}
\end{align}
The self-consistent equation (\ref{q-sc}) is expressed as 
\begin{align}
(\gamma \tilde c+1)^3
=\frac{\tilde c}{c}(\gamma c+1)^3.
\label{app:dp_dc_0_2}
\end{align} 
Thus, the right-hand side of (\ref{app:dp_dc_0_1}) for the
special values of $c$ satisfying (\ref{app:dp_dc_0_2}) is
calculated as
\begin{eqnarray}
&&4\gamma \frac{(\gamma c+1)^3 (c-\tilde c ) }{c (\gamma c^2+\gamma(q-1)\tilde c+1)^2  } \nonumber\\
&&\times \left(1-\frac{\gamma c^2+\gamma (q-1)\tilde c^2 +c+(q-1)\tilde c }{ \gamma c^2+\gamma (q-1)\tilde c^2 +1} \right),
\end{eqnarray}
which turns out to be zero from (\ref{tilde-c}).

\section{Estimation of $\Delta$}
\label{app-sec:latent heat}

In this section, we estimate the value of $\Delta$ defined by
(\ref{Def_Delta}) for the model we study.

\begin{figure}[H]
    \subfigure[]{%
        \includegraphics[clip, width=0.5\columnwidth]{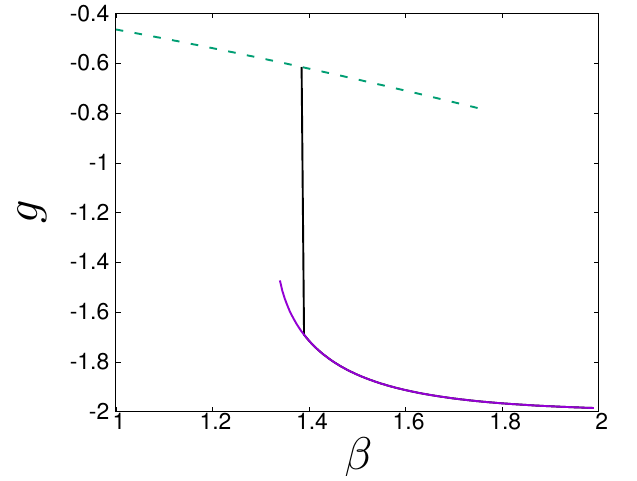}       
        \label{E_hys_rand}
                 }%
    \subfigure[]{%
        \includegraphics[clip, width=0.5\columnwidth]{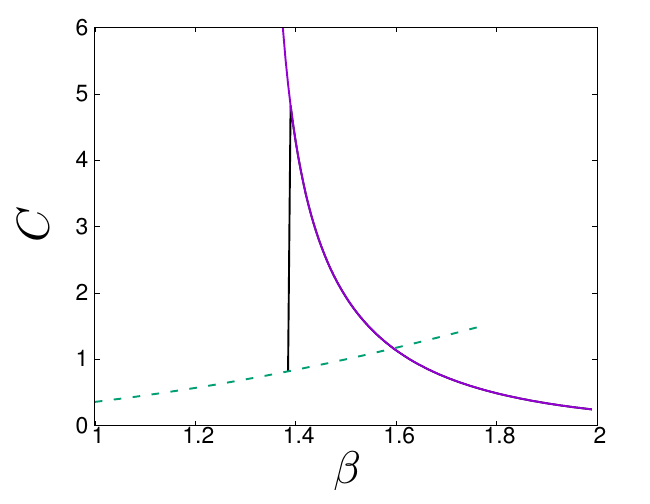}
      \label{heat_capacity_random}
        }%
\caption{ (a) Potential energy density $g$ as a function of
$\beta$. The solid (purple) line represents $g=g_*$ in the range
$\beta > \beta_{\rm sp}$. The dashed (green) line represents $g=g_0$
in the range $\beta < \beta_{\rm un}$. The black line indicates
$\beta=\beta_c$.
(b) Heat capacity per unit volume $C$ as a function of $\beta$.
The styles (colors) of lines correspond to the graphs in (a).}
 \end{figure}

We first calculate
the energy density $h$ defined as
\begin{equation}
h \equiv \lim_{|\Lambda| \to \infty} \frac{1}{|\Lambda|}
\sum_{\sigma,p} P_{\rm can}(\sigma,p) H(\sigma,p),
\end{equation}
where $P_{\rm can}(\sigma,p)$ is given in (\ref{ss-dis}). 
Using the free energy density $f$ calculated in Sec. \ref{SM},
we express the energy density $h$ as
\begin{align}
h=T+g,
\end{align} 
where $g$ is the potential energy density given by
\begin{align}
g(\beta)\equiv\frac{\partial}{\partial \beta} \left( \beta f(\beta) \right).
\end{align} 
As with the free energy density, $g_0(\beta)$ and $g_*(\beta)$
denote the potential energy densities corresponding to the
trivial solution $u_0$ and the nontrivial solution $u_*$ of (\ref{q-sc}),
respectively. In Fig.~\ref{E_hys_rand}, $g_0(\beta)$ and $g_*(\beta)$
are displayed. Then, the latent heat per unit volume $T_c\delta s$
at the equilibrium transition temperature is determined by
the entropy jump defined as
\begin{align} 
\delta s\equiv\beta_c(g_0(\beta_c)-g_*(\beta_c)).
\end{align}
 
For the model with $q=10$, we obtain $T_c\delta s\simeq 1.07$.
Next, we consider the heat capacity per unit volume $C$ expressed as
  \begin{align}
  C(\beta)\equiv \frac{\partial h}{\partial T}
  =1+\frac{\partial g}{\partial T}.  
  \end{align}
Using similar notations, we obtain $C_0(\beta)$ and $C_*(\beta)$ from
$g_0(\beta)$ and $g_*(\beta)$. These are shown at the bottom of 
Fig.~\ref{heat_capacity_random}. For the cases 
$q=10$ and $T_L=1.01T_c$, we obtain $C_*(T_L)\simeq 6.95$. 
Therefore, for the model we numerically study, we have 
\begin{equation}
\Delta\simeq 0.05,
\end{equation}
which is less than unity. 
Note that in the stochastic model studied in this paper, $C$ corresponds to $c_p$ in the phase-field model.

\section{Estimation of $D$}\label{app-sec:Diff}
\begin{figure}[H]
\centering
\includegraphics[width=0.5\textwidth]{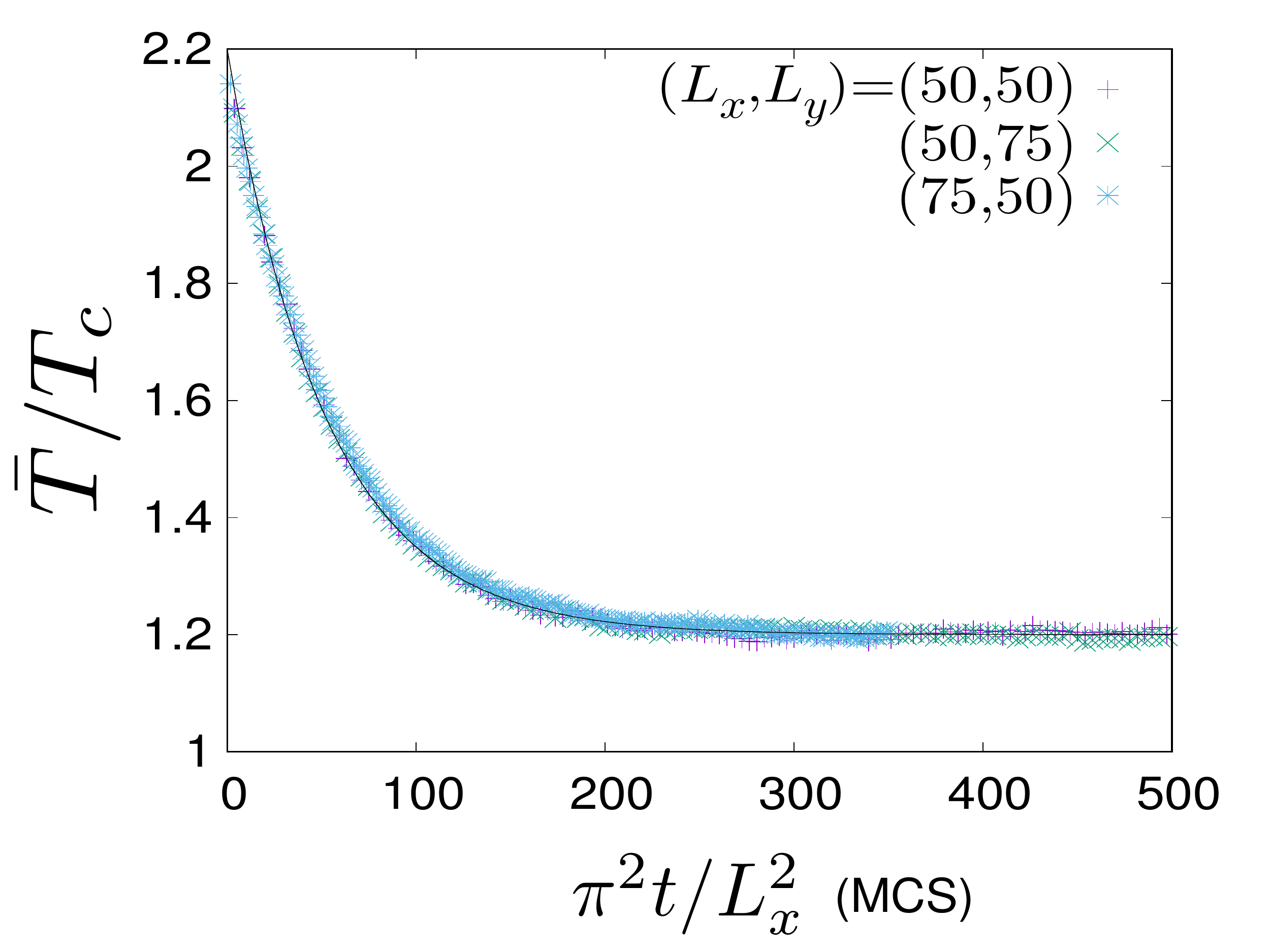}
\caption{$\bar{T}/T_c$ as a function of $\pi^2t/L_x^2$.
The solid line represents the fitting curve (\ref{fit}). 
}
\label{Diffconst}
\end{figure}

In this section, we estimate the value of the thermal
diffusion constant $D$ by measuring the relaxation property
of the temperature profile $T(x,t)$, where
\begin{equation}
T(x,t)\equiv\frac{1}{L_y}\sum_{y=1}^{L_y} p_{x,y}(t).
\end{equation}
For simplicity, we study the case $T_R=T_L=1.2T_c$ with
the initial condition
\begin{equation}
T(x,0)\equiv 1.2T_c+\sin\left(\frac{\pi(x-1)}{L_x-1}\right).
\end{equation}
To realize the initial condition $T(x,0)$, $\sigma_i$ is randomly chosen with equal probability and $p_i=T(i_x,0)$ for any $i$.
We then define the spatial average of the local temperature as
\begin{equation}
\bar{T}(t)\equiv
\left\langle \frac{1}{L_x}\sum_{x=1}^{L_x}T(x,t)  \right\rangle, \label{Tbar}
\end{equation}
where $\langle \cdot \rangle$ denotes the average over ten
independent samples.  Assuming the diffusion equation for $T(x,t)$, we have
\begin{equation}
\frac{\bar{T}(t)}{T_c}=1.2+Be^{-D\pi^2 t/L_x^2},
\label{fit}
\end{equation}
where $D$ is the thermal diffusion constant and $B$ is a parameter associated with the initial condition. As shown in Fig.~\ref{Diffconst}, we find that the fitting of (\ref{Tbar}) with (\ref{fit}) works well 
 for various system sizes with $B = 1$. From this fitting,
we estimate $D = 1.9\times 10^{-2}$.

\input

\end{document}